\pdfoutput=1
\documentclass[
 reprint,
 amsmath,amssymb,amsfonts,
 aps,
 prab,
]{revtex4-2}

\usepackage{array}
\usepackage{graphicx}
\usepackage[allcolors=blue]{hyperref}
\graphicspath{{figures/}}

\begin{document}

\title{Design and Operation of a 3-Crystal Multi-Volume Reflection Array
       for Loss Reduction in CERN SPS Slow Extraction}

\author{F.~M.~Velotti}
\email{francesco.maria.velotti@cern.ch}
\author{M.~Bay}
\author{T.~Calvet}
\author{M.~Calviani}
\author{G.~Daher}
\author{L.~S.~Esposito}
\author{M.~A.~Fraser}
\author{B.~Goddard}
\author{M.~Howes}
\author{V.~Kain}
\author{E.~Matheson}
\author{A.~Perillo~Marcone}
\author{R.~Seidenbinder}
\affiliation{CERN, Geneva, Switzerland}

\date{\today}

\begin{abstract}
The flux of protons slow-extracted from the CERN Super Proton Synchrotron (SPS) to the North Area experiments is limited by the induced radioactivity of the beam lost on the electrostatic septum during third-integer resonant extraction. Crystal shadowing, in which a thin bent silicon crystal deflects the protons that would otherwise hit the septum wires, has been in operation at the SPS since 2021; a single crystal at a non-local position halved the extraction losses relative to the local crystal alone, short of the fourfold reduction that future high-intensity operation requires. In this paper, the design, deployment and operation of a 3-crystal Multi-Volume Reflection Array (MVRA) at the non-local position in the fourth long straight section (LSS4) are presented. The array geometry was designed by Multi-Fidelity Bayesian Optimization (MFBO), combining a multilayer-perceptron surrogate with full particle tracking; it predicts a fourfold loss reduction with three crystals and up to tenfold with an ideally aligned array of four to five. At its optimum the as-built array reaches, and at its best point exceeds, the predicted factor of four, and beam dynamics simulations matched to the measured position and angle scans reproduce their shape to within $0.04$ of the crystals-retracted loss. In physics production an extremum-seeking controller has held the array on its operating plateau, with a median loss reduction by a factor of $3.6$ over $1.14\times 10^{5}$ extraction cycles relative to the pre-shadowing baseline. The remaining margin to the target, the limitations of the present installation and the upgrade to a 4- or 5-crystal array are discussed.
\end{abstract}

\maketitle

\section{Introduction}
\label{sec:intro}

The third-integer resonant slow extraction of the CERN Super Proton Synchrotron (SPS), which delivers $400$~GeV/$c$ protons to the North Area fixed-target programme, deposits a non-negligible fraction of the extracted intensity on the wires of the electrostatic septum (ZS). These losses activate the second long straight section (LSS2), dominate the local radiological budget of the SPS and cap the intensity that can be delivered sustainably.

Future high-intensity programmes at the SPS, in particular the proposed Beam Dump Facility hosting the SHiP experiment, call for up to $4\times 10^{19}$ protons on target per year~\cite{BDF_CDS_2020}, roughly four times the present North Area delivery. Since the activation of the ZS region scales with the integrated beam loss, holding it at its present, already limiting level requires reducing the extraction loss \emph{per proton} by the same factor of four. That target was set in the first crystal-shadowing paper~\cite{Velotti_2019_PRAB} and is the one against which the MVRA is measured here.

The use of bent monocrystals to deflect charged particles via planar channeling was proposed in the mid-1970s~\cite{Tsyganov_1976} and demonstrated for beam extraction at the IHEP Protvino accelerator complex over the following decades~\cite{Afonin_2001,Afonin_2021}. At Fermilab a channeling crystal extracted $900$~GeV protons from the Tevatron at about $25\%$ efficiency during collider operation~\cite{Carrigan_2002}. The SPS has its own crystal-extraction history: a bent silicon crystal extracted a $120$~GeV/$c$ coasting beam from the ring at up to $15\%$ efficiency in the 1990s~\cite{Elsener_1996}, and bent crystals were later studied there for the collimation of high-energy hadron beams~\cite{Scandale_2015_SPS}. An early demonstration with a passive amorphous scatterer~\cite{Goddard_2017} showed that diverting particles away from the ZS wires before extraction reduces the extraction loss, and complementary techniques developed for the SPS slow extraction include a wire diffuser~\cite{Goddard_2020_diffuser} and dynamic-bump-and-octupole schemes~\cite{Fraser_2019}.

Crystal shadowing, and in particular its non-local variant, was first proposed in~\cite{Velotti_Thesis}: a single bent crystal deflects the separatrix particles off the septum wires on their last turns before extraction. In its \emph{local} form the crystal sits in LSS2, immediately upstream of the ZS; this was first demonstrated at the SPS in 2019~\cite{Velotti_2019_PRAB,Velotti_2019_IPAC,Esposito_2019} and consolidated through automated alignment and multi-crystal operation over the following years~\cite{Velotti_2022_IPAC}. In its \emph{non-local} form the crystal sits in LSS4, several straight sections upstream, and its deflection is mapped onto the septum aperture through the betatron phase advance and amplified by the non-linear extraction optics between LSS4 and LSS2; this was demonstrated in 2022~\cite{Velotti_2023_IPAC} and consolidated in 2024~\cite{Velotti_2024_IPAC}, when it halved the extracted loss per proton. Both configurations are reviewed in Section~\ref{sec:concepts}; the non-local single crystal is the baseline on which the MVRA is built.

The non-local configuration could not at first be sustained beyond a few hours: magnetic non-reproducibility and hysteresis in the SPS main magnets drift the closed orbit, and the single-crystal shadowing dip is too narrow for an open-loop goniometer setting to stay inside it (Section~\ref{sec:concepts:nonlocal}). The present work removes both obstacles: the array of Section~\ref{sec:concepts:mvra} widens the working window by an order of magnitude, and the controller of Section~\ref{sec:feedback} holds the array inside it through routine production.

Two gaps had to be closed to get there. First, the factor of two of a single non-local crystal is relative to the machine with the local crystal in, which leaves it well short of the factor of four that the activation budget requires relative to the pre-shadowing machine, and going further means stacking the deflections of several bent crystals. Second, the resulting geometry, with per-crystal position and width, inter-crystal angular stagger and number of crystals all open at once, has no cost-aware design method in the literature; earlier multi-crystal arrays were hand-tuned or optimised one parameter at a time.

This paper follows the array from design to operation. Section~\ref{sec:concepts} introduces the MVRA, a sequence of bent crystals in volume reflection whose deflections add, and the shadowing concepts behind it. Section~\ref{sec:mvra_design} describes its design by a Multi-Fidelity Bayesian Optimization (MFBO) campaign that combines a fast multilayer-perceptron surrogate with full particle tracking under a cost-aware multi-fidelity utility. Sections~\ref{sec:setup} and~\ref{sec:expdata} give the experimental setup and the measurements, compared with Xsuite tracking matched to the machine (sim/machine matching). Section~\ref{sec:feedback} reports the extremum-seeking controller that has held the array at its operating point through routine production in 2026. Section~\ref{sec:activation} gives the loss budget against the no-crystal machine and the single-crystal configurations, with the limitations, and Section~\ref{sec:applicability} discusses other machines. The surrogate-guided and model-based controllers developed since 2019 for the non-local (TECA-single) single crystal will be reported separately.

\section{MVRA and Crystal Shadowing Concepts}
\label{sec:concepts}

A bent silicon crystal deflects a proton coherently in one of two ways, selected by the angle between its trajectory and the bent atomic planes~\cite{Biryukov_Book}. Within the Lindhard critical angle~\cite{Lindhard_1965}, $\theta_c = 10.6~\mu$rad for $400$~GeV/$c$ protons in Si(110), the proton is \emph{channeled} between the planes and follows the full bend. At larger angles, as long as the trajectory still crosses the bent planes somewhere inside the crystal, it is \emph{volume-reflected} (VR): it receives a small kick of sign opposite to the bend. Volume reflection was predicted in 1987~\cite{Taratin_1987}, first observed with protons in 2006~\cite{Ivanov_2006}, and measured at $400$~GeV with an efficiency above $95\%$~\cite{Scandale_2007_VR} and then $98\%$ per plate~\cite{Scandale_2008_doubleVR}. It occurs over the whole angular range swept by the bend, so its acceptance is the bending angle, $160~\mu$rad for the strips used here, against a channeling window of order $\theta_c$.

The kick is small and set by how tightly the crystal is bent. A fit to silicon and germanium data from $1$ to $400$~GeV gives $\theta_\mathrm{VR} = (1.03 \pm 0.09)\,\theta_c \log_{10}(R/R_c)$ for $1 < R/R_c < 30$, where $R_c$ is the critical radius below which channeling is lost~\cite{Biryukov_2016_VR,Scandale_2008_curvature}. With $R_c = 0.68$~m and the $2$~mm strips bent by $160~\mu$rad ($R = 12.5$~m, $R/R_c = 18$), a single traversal gives $\theta_\mathrm{VR} = 13.8 \pm 1.2~\mu$rad and three in sequence $41 \pm 4~\mu$rad. The simulated median deflection of the volume-reflected protons at the operating point, $+37~\mu$rad (Section~\ref{sec:expdata:matching}), is about $10\%$ below that sum. A single strip reflects about $98\%$ of the protons that traverse it, but a proton receives the full cumulative kick only if it lies inside the acceptance of every strip in turn, and that fraction was measured at $75$ to $82\%$ in the test beam~\cite{Aberle_2026_Channeling}. The compounding sets the ceiling on how many strips are worth adding.

\subsection{Single-crystal shadowing}
\label{sec:concepts:nonlocal}

\begin{figure}
\includegraphics[width=\columnwidth]{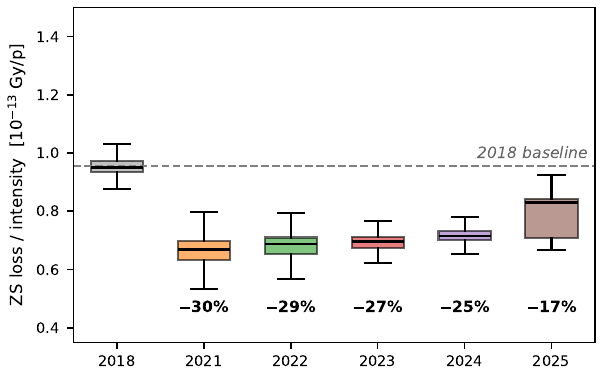}
\caption{Intensity-normalised ZS loss (sum of the six beam loss monitors (BLMs) of the five ZS tanks and of the TCE at the extraction-channel exit, per extracted proton) in operation across the years of local-shadowing deployment. The dashed line is the 2018 pre-shadowing baseline; boxes show the per-year distribution with the local crystal (TECS) in operation, annotated with the reduction of the annual mean relative to 2018. The reduction is $17$--$30\%$ across 2021--2025, $\sim 25\%$ on average.}
\label{fig:local_losses}
\end{figure}

The local crystal (TECS) sits a few metres upstream of the ZS, aligned with its wires, and deflects into the septum gap the protons that would otherwise strike the wires on their final turn; the phase-space picture is given in~\cite{Velotti_2019_PRAB}. It sustains a loss reduction of about $25\%$~\cite{Velotti_2024_IPAC}, shown year by year in Figure~\ref{fig:local_losses}. Full channeling would reach $\sim 40\%$, but the channeled beamlet is then deflected too far and is lost on the aperture of the TT20 transfer line, so the crystal is operated in volume reflection with the shallower reduction.

The non-local crystal (TECA-single) in LSS4 halves the loss relative to the machine with only the local crystal in~\cite{Velotti_2024_IPAC}, and that factor of two is the baseline throughout this paper. Its working point is the transition between volume reflection and amorphous scattering, where the shadowing dip is only $\sim 20~\mu$rad wide against a $\sim 10~\mu$rad goniometer step, too narrow for an open-loop setting to be held inside it once the extraction optics drift~\cite{Velotti_2024_IPAC}.

Two earlier numbers for the single non-local crystal have since been corrected. The tracking study of~\cite{Velotti_2019_PRAB} predicted about a factor of four because of an error in the tracking model; the campaigns of~\cite{Velotti_2023_IPAC,Velotti_2024_IPAC} measured the factor of two used here. And an angular scan on a $25~\mu$rad grid stepped over a feature roughly $15~\mu$rad wide and returned the shoulder beside it, so a published reduction was low by some fifteen percentage points; this is why the angle scans of Section~\ref{sec:expdata:match} are finely stepped and why the estimator was chosen by measurement (Section~\ref{sec:expdata:method}).

\subsection{From single-crystal shadowing to the MVRA}
\label{sec:concepts:mvra}

The Multi-Volume Reflection Array replaces the single crystal with a sequence of bent strips, each in volume reflection, whose kicks add coherently; the design campaign of Section~\ref{sec:mvra_design} finds the reduction scaling toward $\sim 10\times$ for four to five strips. The array is built to add kicks; its wide acceptance keeps it inside its working window as the machine drifts (Section~\ref{sec:feedback}).

Stacked volume reflection has a test-beam history. Two plates at $400$~GeV/$c$ gave $23.23 \pm 0.18~\mu$rad, twice the single-crystal value, at above $97\%$ efficiency each~\cite{Scandale_2008_doubleVR}; five aligned quasimosaic plates gave $52.96 \pm 0.14~\mu$rad, consistent with the sum of their individually measured reflections, at above $80\%$ efficiency over a $70~\mu$rad interval of array orientation~\cite{Scandale_2009_MVR}; and eleven of fourteen bent strips in the SPS extracted beam reflected coherently for $110~\mu$rad at $88\%$~\cite{Scandale_2010_MVR14}. The same stacking has been obtained inside a single crystal entered at a small angle to a crystallographic axis, so that successive sets of inclined planes reflect in turn~\cite{Tikhomirov_2007}. The per-plate angles of order $10~\mu$rad in that work are the scale of the present array and an external check on Section~\ref{sec:mvra_design}. What is new here is the application: such an array inside a circular machine, on the separatrix of a running third-integer extraction, shadowing an electrostatic septum and held there by feedback through physics production.

\begin{figure*}
\includegraphics[width=\textwidth]{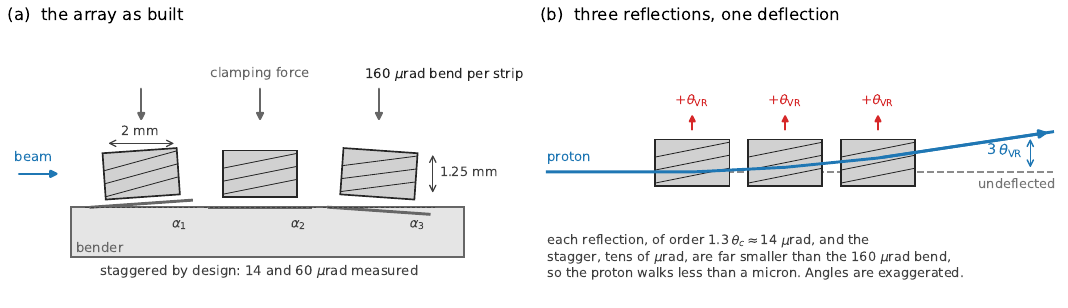}
\caption{The 3-crystal Multi-Volume Reflection Array. \emph{(a)} The array as built, in plan view: three bent silicon strips clamped on a common bender, each seated on its own contact face, so that the strip-to-strip angles $\alpha_i$ are fixed at assembly and the goniometer then rotates the array as a whole. Each strip presents $1.25$~mm to the beam across the extraction plane and $2$~mm of silicon along it, and is bent by $160~\mu$rad; the strips are deliberately staggered in angle, by a measured $14$ and $60~\mu$rad on the as-built assembly~\cite{Aberle_2026_Channeling} (Section~\ref{sec:expdata:match}). The strip cross-section is drawn to scale, the tilts and the gaps between strips are not. \emph{(b)} Why three strips outperform one: a proton is volume-reflected once in each strip, and the same-sign kicks add to $\approx 3\,\theta_\mathrm{VR}$ while the acceptance stays the full angle swept by the bend. With $\theta_\mathrm{VR} \approx 14~\mu$rad the proton walks less than a micron across the array; the angles are exaggerated by orders of magnitude, and the accumulated deflection only becomes visible after a drift.}
\label{fig:array}
\end{figure*}

Figure~\ref{fig:array} shows the array as built. Its strips are deliberately \emph{staggered} in angle~\cite{Aberle_2026_Channeling}, so that the acceptance of each overlaps the beam already deflected by the strips before it and the full cumulative kick is delivered over a wider range of incoming angle; the matched per-crystal offsets of Section~\ref{sec:expdata:match} are read against this designed stagger. Strips, bender and assembly come from the DECRYCE crystal-production programme at CERN, with an autocollimator validation of the assembly and a beam-based validation on a silicon-pixel telescope~\cite{Esposito_2025_SX,Aberle_2026_Channeling}. The array is mounted on the TECA goniometer in LSS4, which carried the non-local single crystal (TECA-single, identical in geometry to TECS) for the 2022 to 2024 demonstrations; the two are alternative payloads on one mount.

\section{MFBO Design of the MVRA}
\label{sec:mvra_design}

\subsection{Design problem}
\label{sec:mvra_design:problem}

The design problem is to choose the geometry of an array of bent silicon crystals operated in volume reflection so as to maximise the loss reduction at the ZS. Four design parameters describe the array (Table~\ref{tab:mfbo_dof}): the width of each crystal, the angular stagger between successive crystals, the number of crystals, and the sign of the volume-reflection deflection, which selects whether the array deflects the separatrix particles towards the machine inside or the machine outside. Two further quantities, the shadowing angle of the array centre and its position relative to the ZS aperture, are operating setpoints. Every geometry has its own optimum pair, found in the machine by scanning (Section~\ref{sec:expdata}), so the optimisation searches them jointly with the design parameters and reports the optimum setpoint with each geometry.

\begin{table}[h]
\squeezetable
\centering
\caption{The four design parameters of the array and the two operating setpoints, all searched jointly by the MFBO campaign.}
\label{tab:mfbo_dof}
\begin{tabular}{lll}
\hline
Design parameter & Range & Type \\
\hline
Crystal width & $0.8$ to $1.9$~mm & continuous \\
Inter-crystal stagger & $-20$ to $+21~\mu$rad & continuous \\
Number of crystals $n_\mathrm{cry}$ & 1 to 5 & discrete \\
VR sign direction & $\pm 1$ & discrete \\
\hline
Operating setpoint & Range & Type \\
\hline
Array shadowing angle & $-1000$ to $-400~\mu$rad & continuous \\
Position relative to ZS & $-16.2$ to $-15.5$~mm & continuous \\
\hline
\end{tabular}
\end{table}

The continuous ranges in Table~\ref{tab:mfbo_dof} bracket the operating envelope of the LSS4 goniometer and of the SPS extraction geometry. The objective is the relative beam loss at the ZS aperture, normalised to the \emph{crystals-retracted} loss for the same incoming distribution: the reference run moves every crystal a metre clear of the beam and sets its deflection to zero; the ZS aperture stays in place, so the denominator is bare extraction.

The cost structure is asymmetric. A high-fidelity evaluation tracks $5\times 10^4$ particles for $300$ turns through the SPS lattice, represented sector by sector by Taylor maps generated with MAD-X PTC~\cite{maptrack_sw}, with the array and the ZS modelled by the crystal and septum routines of pycollimate~\cite{pycollimate_sw}, and takes tens of minutes on CPU; a low-fidelity evaluation by a pre-trained multilayer-perceptron surrogate takes milliseconds. A grid search of the six-dimensional space at high fidelity is not feasible, and a cost-aware multi-fidelity optimisation is the alternative. The design model predates and is independent of the Xsuite model of Section~\ref{sec:expdata}, and the design prediction is tested against the measurement, never against the Xsuite model.

\subsection{Multi-fidelity Bayesian optimisation}
\label{sec:mvra_design:mfbo}

The optimisation uses a multi-fidelity Gaussian-process (GP) model with a multi-fidelity knowledge-gradient acquisition function~\cite{Wu_Frazier_2017,Wu_2020_UAI}, weighted by an affine fidelity-cost model so that each evaluation point is chosen jointly with the fidelity level that maximises information gain per unit cost. The low-fidelity model is the multilayer-perceptron surrogate, pre-trained on $500$ samples of the same tracking model from a prior crystal-shadowing dataset (a transfer-learning warm start in the spirit of~\cite{Feurer_2018_RGPE}); the high-fidelity model is the full tracking simulation. The two levels are encoded by an auxiliary input of value $0.4$ (low) or $1.0$ (high), and the cost model, with no fixed cost, charges each evaluation in proportion to this input. The initial GP training set consisted of 100 high-fidelity samples from the prior dataset and 100 low-fidelity samples drawn uniformly across the design space; the BO loop then ran for 50 iterations. The campaign was run twice with this set-up, and the GP of Figure~\ref{fig:mfbo_surface} is the one from the second run; the fidelity mix of its acquisitions is given in Appendix~\ref{app:mfbo}. The dependence on the number of crystals was checked after the loop by direct high-fidelity runs at integer $n_\mathrm{cry}$ for VR towards the inside, the blue markers of Figure~\ref{fig:mfbo_surface}(b). Bayesian optimisation of accelerator systems is reviewed in~\cite{Roussel_2024}, with applications to on-line tuning~\cite{Duris_2020} and high-dimensional, multi-objective and learning-based variants developed in~\cite{Kirschner_2019,Roussel_2021,Kaiser_2024}.

\subsection{Optimisation results}
\label{sec:mvra_design:results}

\begin{figure}
\includegraphics[width=\columnwidth]{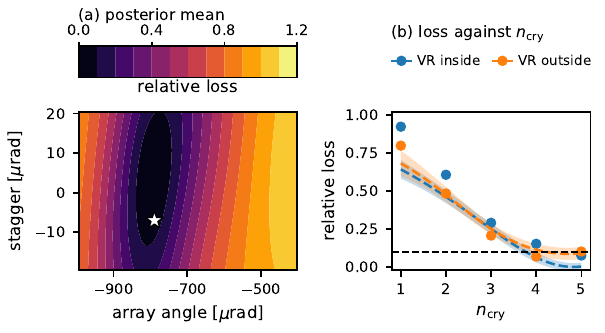}
\caption{The MFBO design surface. \emph{(a)} Gaussian-process posterior mean of the relative-loss objective in the (array shadowing angle, inter-crystal angular stagger) slice through the six-dimensional search space, with the other coordinates at the optimum; the star marks the optimum adopted in the specification, an angle of $-789~\mu$rad and a stagger of $-7~\mu$rad. \emph{(b)} Relative beam loss at the ZS aperture against the number of crystals, with the other search coordinates at their optimum: GP posterior mean (dashed line) and one-standard-deviation band for the two VR sign directions, with markers at integer $n_\mathrm{cry}$. The markers for VR towards the inside (blue) are direct high-fidelity tracking runs; those for VR towards the outside (orange) are predictions of the low-fidelity surrogate, not tracking. The horizontal dashed line at $0.1$ marks the factor-of-ten level relative to the crystals-retracted baseline. At four and five crystals only a few tracked particles are still lost at the ZS, so the tracking markers there carry a larger statistical uncertainty than at fewer crystals. The GP has no high-fidelity training point at $n_\mathrm{cry} = 1$ (the prior dataset covers two to five crystals), so its band there is the surrogate's extrapolation and both markers sit above it; at $n_\mathrm{cry} = 3$, the array built, both markers agree with the posterior to within $1.3$ standard deviations.}
\label{fig:mfbo_surface}
\end{figure}

Figure~\ref{fig:mfbo_surface}(a) shows the GP posterior at the optimum in the (array shadowing angle, inter-crystal stagger) plane. The basin is extended and comparatively flat: changes of the stagger of order tens of microradians move the predicted loss only weakly, so the optimum is robust against alignment tolerances of that size. The posterior standard deviation, read where the mean lies within $0.05$ of its minimum, is $0.02$ at the optimum and $0.04$ at the edge of the basin, so the posterior is well constrained across the basin; the flatness is a property of the GP surface and was not checked point by point with tracking.

The load-bearing design result is Figure~\ref{fig:mfbo_surface}(b), the optimised relative loss against the number of crystals. A single crystal in volume reflection gives only a modest reduction, to $0.92$ in tracking for the inside direction and $0.80$ in the surrogate for the outside one, the effect of a single volume-reflection kick; the factor of two of~\cite{Velotti_2024_IPAC} comes from the narrower working point at the volume-reflection-to-amorphous transition (Section~\ref{sec:concepts:nonlocal}). The 3-crystal array reaches about $0.25$, a factor of four, and the GP estimate reaches the factor-of-ten level at $n_\mathrm{cry} = 4$--$5$. That estimate carries a larger error bar than the one at three crystals: the high-fidelity data there are sparse, and the direct tracking runs at four and five crystals, which give $0.15$ and $0.08$ (factors of $6.5$ and $13$) and sit above the GP mean, rest on only a few lost particles, as the particle number was not raised to stabilise them.

The factor of ten is a projection for an \emph{ideal} array. The surface carries no manufacturing tolerances in its noise model (Appendix~\ref{app:mfbo}), so each additional crystal is assumed to sit inside the volume-reflection acceptance of the array as a whole. That assumption is the one the as-built hardware strains: the matched configuration of Section~\ref{sec:expdata:match} needs effective per-strip angular offsets spanning $62.5~\mu$rad on three crystals, and the largest array operated at this momentum reflected coherently from eleven of its fourteen strips (Section~\ref{sec:concepts:mvra}). Whether a 4- or 5-crystal array reaches the projection will be decided by how many of its strips can be aligned.

\subsection{As-built configuration}
\label{sec:mvra_design:limitations}

The array has three crystals for a hardware reason. It was built with four strips, but metrology and a test-beam scan showed one of them tilted some $300~\mu$rad from the other three, too far to reflect coherently with them, and it was removed (Section~\ref{sec:expdata:match}). A bender of this design can hold more, so the 3-crystal MVRA targets the factor-of-four point of Figure~\ref{fig:mfbo_surface}(b) and leaves the factor of ten to a 4- or 5-crystal upgrade with better per-crystal alignment. Table~\ref{tab:mvra_optimum} sets the functional specification~\cite{Fraser_2025_FS} against what was built. The specification asked for four strips, one fewer than the optimum's five, with a stagger of $-7~\mu$rad per strip and a $\pm 20~\mu$rad tolerance. The array as built departs from it: its larger stagger step, $60~\mu$rad, lies outside the range the optimisation explored (Table~\ref{tab:mfbo_dof}), and its bend is $160~\mu$rad against the $174~\mu$rad the specification assumed. The comparison of Section~\ref{sec:expdata:meas} with the design prediction is therefore between the design point and a neighbouring device, not a test of the same geometry.

\begin{table}[h]
\squeezetable
\centering
\caption{The array as specified and as built. The specification is the released functional requirement of the array~\cite{Fraser_2025_FS}, with the miscut from the SPS crystal-manufacturing specification; the bending angle (a) is the value the specification's own simulations assumed, with no requirement placed on it. The as-built stagger is the autocollimator metrology~\cite{Aberle_2026_Channeling}, the transverse offsets and the deflection come from the matched model (Section~\ref{sec:expdata}), and the built acceptance is the full angular range of the measured angle scan over which the reduction relative to the crystals-retracted loss exceeds the design's factor of four (Section~\ref{sec:expdata:meas}). The specified minimum of $\pm 50~\mu$rad is a full range of $100~\mu$rad, with no loss criterion attached. Blank cells were not measured on these strips.}
\label{tab:mvra_optimum}
\begin{tabular}{lcc}
\hline
Parameter & Specified & Built \\
\hline
Strip length & $\leq 2$~mm & $2$~mm \\
Crystal width & $1.25 \pm 0.10$~mm & $1.25$~mm \\
Bending angle$^{\mathrm{a}}$ & $174 \pm 1~\mu$rad & $160~\mu$rad \\
Torsion & $\leq 5~\mu$rad/mm & \\
Miscut & $0 \pm 200~\mu$rad & \\
Crystals & $4$ & $3$ \\
Stagger & $-7(n{-}1) \pm 20~\mu$rad & $14$, $60~\mu$rad \\
Transverse offsets & $\pm 0.1$~mm & $-35$, $-75$, $+15~\mu$m \\
VR deflection & $\approx 52~\mu$rad & $+37~\mu$rad (sim.) \\
VR acceptance & $\geq \pm 50~\mu$rad & $105~\mu$rad ($>4\times$) \\
\hline
\end{tabular}
\end{table}

\section{Experimental Setup at the SPS}
\label{sec:setup}

\subsection{Diagnostics}
\label{sec:setup:diagnostics}

LSS2 contains the local crystal TECS and the ZS, and LSS4 contains the TECA goniometer carrying the MVRA. The loss observable used throughout is built from six beam loss monitors (BLMs) in the ZS region: five (ZS1 to ZS5) at the corresponding ZS tank segments and a sixth (TCE) at the thin-walled extraction-channel exit immediately downstream. A further set of BLMs in LSS4, beside the TECA goniometer, reads the losses the array itself generates by scattering and inelastic interaction, and is used to verify that the loss removed from the ZS is not offset by loss at LSS4. The extracted intensity per cycle comes from the dedicated SPS intensity diagnostic, and beam position monitors (BPMs) in LSS2 (six, both planes) and LSS4 (five, both planes) provide the closed-orbit context used in the sim/machine matching of Section~\ref{sec:expdata}; the controller of Section~\ref{sec:feedback} does not use them.

Both goniometers expose a position and an angle setpoint of the crystal mount relative to the local beam axis through the SPS control system, quantised at $10~\mu$m and $10~\mu$rad; sub-quantisation move requests are rejected by the hardware.

\subsection{Intensity-normalised loss observable}
\label{sec:setup:obs}

The central observable of the paper is an intensity-normalised total loss at the ZS region, defined as

\begin{equation}
\mathrm{tot\_loss} = \frac{\sum_{i \in \{Z\!S1,\ldots, Z\!S5,\,T\!C\!E\}} \mathrm{BLM}_i}{I_\mathrm{ext}\, L_\mathrm{nom}},
\label{eq:tot_loss}
\end{equation}

where the sum runs over the six BLMs above, $I_\mathrm{ext}$ is the extracted beam intensity per cycle, and $L_\mathrm{nom} = 0.95\times 10^{-13}$~Gy/proton is a reference dose-per-proton constant chosen to keep $\mathrm{tot\_loss}$ of order unity for the pre-shadowing machine. Cycles with $I_\mathrm{ext} < 10^{12}$ protons (partial or aborted spills) are filtered out. The same observable underlies the multi-year record of the crystal-shadowing programme, on which Section~\ref{sec:activation} compares the MVRA with the earlier configurations.

A loss reduction is a ratio, and the paper uses four denominators that differ from one another by as much as some of the effects reported (Table~\ref{tab:denominators}). The \emph{crystals-retracted reference}, also written \emph{crystal-out} or \emph{no-crystal}, is the machine with every shadowing crystal clear of the beam and the ZS aperture in place; it is the reference against which design prediction and measurement compare directly. The \emph{amorphous-orientation (AM) baseline} divides a scan by its own loss far from the coherent alignment, where the inserted array acts only as an amorphous scatterer; it differs from the crystals-retracted reference by however much the array shadows before it is aligned at all, about a fifth on the measured angle scan. The \emph{pre-shadowing operational baseline} is the 2018 machine, before any crystal was installed, against which the factor-of-four target was set. The no-crystal machine after the second long shutdown reads $0.93$ of it on the same monitors, both from $55$ both-retracted cycles pooled over 2024 and 2025 and from the retracted end of the 2026 position scan of Section~\ref{sec:expdata:meas}. Split by window, the pooled cycles give $0.93$ (2024, $48$ cycles) and $0.90$ (2025, seven cycles), so we take $\pm 0.03$ as the uncertainty of the factor; it moves the crystals-retracted values quoted in this paper by about $\pm 0.01$. We attribute the difference to the recalibration of the loss monitors and the realignment of the extraction over the shutdown. A loss relative to 2018 divided by $0.93$ is therefore its value relative to the crystals-retracted reference, and the two are not otherwise mixed. The \emph{TECS-only level} is used once, for the within-2026 control of Section~\ref{sec:activation:budget}, because it is free of that recalibration. The constant $L_\mathrm{nom}$ cancels in every ratio.

\begin{table*}[t]
\squeezetable
\centering
\caption{The four denominators of the loss ratios in this paper (Section~\ref{sec:setup:obs}). A loss relative to 2018 divided by $0.93$ is its value relative to the crystals-retracted reference.}
\label{tab:denominators}
\begin{tabular}{@{}>{\raggedright\arraybackslash}p{0.24\textwidth}@{\hspace{8pt}}>{\raggedright\arraybackslash}p{0.33\textwidth}@{\hspace{8pt}}>{\raggedright\arraybackslash}p{0.36\textwidth}@{}}
\hline
Denominator (also written) & The loss is divided by & Used for \\
\hline
Crystals-retracted reference (crystal-out, no-crystal) & the loss with every crystal clear of the beam and the ZS aperture in place & the MFBO objective (Section~\ref{sec:mvra_design}); the simulated scans and the match (Section~\ref{sec:expdata}); the loss budget (Section~\ref{sec:activation:budget}); the design's axis \\
Amorphous-orientation (AM) baseline & the same scan's own loss far from the coherent alignment, where the inserted array acts only as an amorphous scatterer & the matching metric (Section~\ref{sec:expdata:method}); about $0.8$ of the crystals-retracted reference on the measured angle scan \\
Pre-shadowing operational baseline (2018) & the 2018 machine, no crystal installed & Figure~\ref{fig:local_losses}; the factor-of-four target; the operational record (Section~\ref{sec:feedback:deployment}) \\
TECS-only level & the 2026 machine with only the local crystal in & the within-2026 control (Section~\ref{sec:activation:budget}) \\
\hline
\end{tabular}
\end{table*}

\subsection{2026 measurement campaign}
\label{sec:setup:campaign}

The 3-crystal MVRA was characterised in 2026 in dedicated machine-development (MD) sessions, which provided the position and angle scans of Section~\ref{sec:expdata}, followed by an extended run on the production slow-extraction beam with the extremum-seeking controller of Section~\ref{sec:feedback} enabled. The scans and the two-dimensional map were all taken on the morning of 6 May 2026. The per-cycle BLM, BPM, intensity and crystal-setpoint data acquired throughout are the dataset of Sections~\ref{sec:expdata} and~\ref{sec:feedback}.

\section{Experimental Data and Simulation/Machine Matching}
\label{sec:expdata}

\subsection{Measurements}
\label{sec:expdata:meas}

The 3-crystal MVRA was characterised by two one-dimensional scans of the loss observable Eq.~\eqref{eq:tot_loss}. The \emph{angle scan} sweeps the array angle through the volume-reflection alignment at a fixed jaw position: while the crystals are within their VR acceptance the separatrix particles are deflected into the ZS gap, clear of the wires, and $\mathrm{tot\_loss}$ drops into a deep dip whose width is set by the volume-reflection-to-amorphous (VR-to-AM) transition. The \emph{jaw (position) scan} sweeps the jaw position with the array at its machine-reference angle, some $2.5$~mrad from the VR alignment, so it probes the array as an amorphous scatterer and produces a much shallower dip. Both are shown with the matched simulation in Section~\ref{sec:expdata:match}.

\begin{figure*}
\centering
\includegraphics[width=0.92\textwidth]{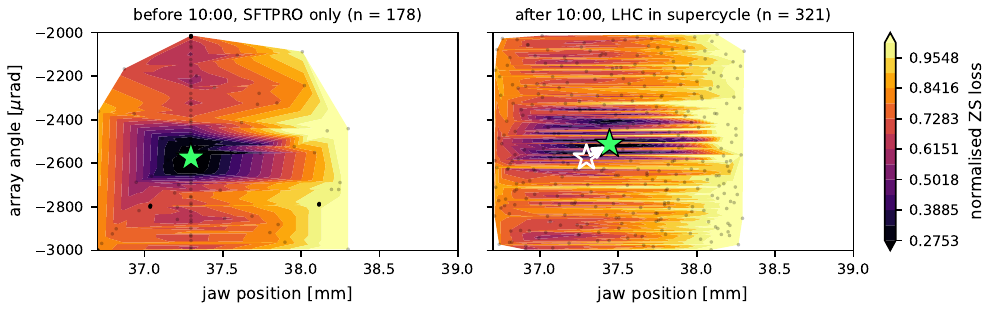}
\caption{Measured intensity-normalised loss Eq.~\eqref{eq:tot_loss} across a two-dimensional Sobol scan of the MVRA jaw position and array angle, split into the two machine periods of the 2026 MD scan: before (left, 178 cycles, SFTPRO-only supercycle) and after (right, 321 cycles) the LHC cycle entered the supercycle at 10:00 local time. Stars mark the loss optimum of each period; between the two the optimum shifts by $+0.15$~mm / $+63~\mu$rad (the open star in the right panel is the first-period optimum).}
\label{fig:sobol_contour}
\end{figure*}

Each dip is characterised by its \emph{depth}, the fractional loss reduction at the minimum relative to the local AM baseline, $\mathrm{depth} = 1 - L_\mathrm{min}/L_\mathrm{AM}$, and its full width at half maximum (FWHM). On the angle scan the loss at the volume-reflection alignment falls to $0.24$ of the AM baseline, a depth of $0.76$ or a factor of $\sim 4$, over an angular FWHM of $\approx 211~\mu$rad, far wider than the $\sim 10~\mu$rad channeling critical angle and consistent with the $160~\mu$rad swept by each bend plus the inter-crystal stagger (Section~\ref{sec:concepts}). The amorphous-orientation jaw scan reaches $0.81$ of the no-crystal level (depth $0.19$) over a spatial FWHM of $1.276$~mm. Relative to the crystals-retracted loss (Table~\ref{tab:denominators}) the angle dip bottoms at $0.184$, the number to set against the MFBO prediction of $0.25$ for $n_\mathrm{cry} = 3$ (Figure~\ref{fig:mfbo_surface}(b)): the as-built array reaches the reduction its design predicted. Each point of the scan is a single extraction cycle; the three lowest, at adjacent settings within $27~\mu$rad, read $0.184$, $0.185$ and $0.188$, and the three retracted cycles that define the reference agree to within $0.3\%$, so the minimum does not rest on one cycle. On that reference the reduction exceeds $4\times$ over $105~\mu$rad, $3.5\times$ over $120~\mu$rad and $3\times$ over $168~\mu$rad, by linear interpolation between the $20~\mu$rad scan steps; the first is the built acceptance of Table~\ref{tab:mvra_optimum}, against a specified full range of $100~\mu$rad. The contrast between a $211~\mu$rad angular acceptance and a millimetre-scale jaw tolerance, visible as the elongated basin of Figure~\ref{fig:sobol_contour}, is the anisotropy that the per-dimension scaling of the feedback controller answers (Section~\ref{sec:feedback}).

The two scans are cuts through a two-dimensional loss surface, which was also mapped directly. Figure~\ref{fig:sobol_contour} shows a quasi-random (Sobol) scan of array angle and position, the measured loss interpolated across the operating box and split into the two machine periods of the scan. In the first period the minimum forms a single, well-defined basin. When the LHC cycle entered the SPS supercycle partway through the scan, the whole basin shifted by $+0.15$~mm in position and $+63~\mu$rad in angle, with a $\sim 123~\mu$m orbit shift at the crystal: an in-situ observation of the magnetic-history sensitivity the controller of Section~\ref{sec:feedback} exists for. The basin stays single and smooth in both periods, which is why a model-free controller is adequate.

\subsection{Simulation framework}
\label{sec:expdata:matching}

The simulation counterpart to the measurements is an Xsuite tracking model of the SPS slow extraction with the array at the TECA s-location, the crystal physics handled by the \texttt{xcoll} collimation module~\cite{Iadarola_2023,VanderVeken_2024}, whose crystal routine derives from the model developed for CERN collimation studies in~\cite{Mirarchi_2015}. Other frameworks for crystal-assisted extraction and collimation exist for the SPS~\cite{Mirza_2015} and on a Geant4-based toolkit~\cite{Sytov_2024}. The model takes the as-built geometry of Section~\ref{sec:mvra_design:limitations} and the lattice with the as-operated 2026 extraction-bump optics. The ZS is a perfect absorber of an \emph{effective} interception width, one of the matched parameters (Appendix~\ref{app:simmodel}). The initial distribution is \emph{captured} by tracking the beam through the full lattice without the crystal and recording every particle that reaches the TECA location; the recorded population is then \emph{replayed} through a few turns with the array in place. The septum idealisation, the pipeline and its validation are described in Appendix~\ref{app:simmodel}.

Figure~\ref{fig:extbeam} shows the extracted beam at the ZS entrance in the matched configuration, from the $356$-capture ensemble of Appendix~\ref{app:budget} replayed at the operating point with the array in and retracted. Of the $1.02\times 10^{7}$ protons extracted with the array in, $80.12\%$ never meet a crystal, $19.59\%$ are volume-reflected (median deflection $+37~\mu$rad), $0.15\%$ are channeled, $0.12\%$ are volume-captured and $0.01\%$ cross a strip with incoherent scattering only. The shadowing is carried by the $1.4\%$ ($1.45\times 10^{5}$ protons of the ensemble, or $1.9\times 10^{11}$ per spill at the measured intensity) that are extracted only with the array in and strike the wires without it. Volume reflection accounts for $99.4\%$ of them; they land at a median $x = 78.0$~mm, one displacement step outside the wire band at $68$~mm and inside the extracted band (first-to-99th-percentile extent $13.2$~mm), so they form no distinguishable satellite. Nearly all protons with a channeling record also carry a volume-reflection record: $61\%$ are deflected along the bend, by $105~\mu$rad at the median and $90$ to $120~\mu$rad for nine in ten, and folded back to $x = 70.1$~mm inside the high-field gap, $37\%$ keep a net kick of the volume-reflection sign, and the remaining $2\%$ end with a kick in between. The channeled fraction, $0.15\%$ on the operating plateau and $0.04\%$ at the loss minimum, is set by the matched per-crystal alignment.

\begin{figure*}
\includegraphics[width=\textwidth]{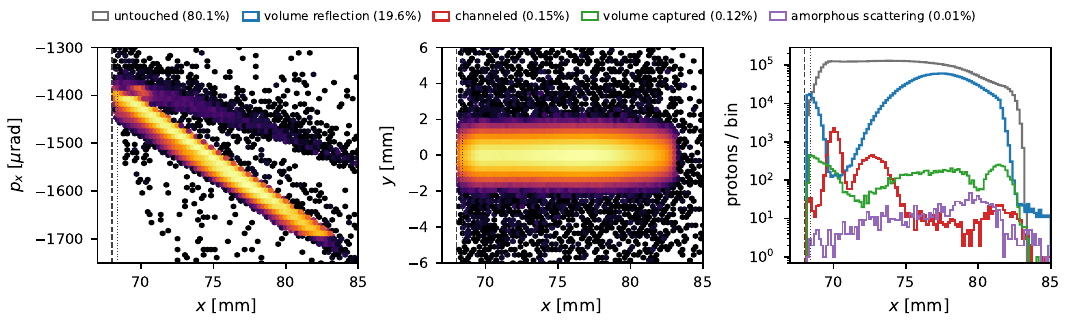}
\caption{Simulated extracted beam at the ZS entrance at the matched configuration of Table~\ref{tab:bestmatch}, at the operating angle ($1.02\times 10^{7}$ extracted protons from $356$ independent captures): transverse phase space $(x, p_x)$ (left) and transverse plane $(x, y)$ (centre), both on a logarithmic colour scale, and the horizontal profile split by crystal process (right). The dashed and dotted lines mark the first ZS wire at $x = 68$~mm and the far side of the wire band. The separatrix arrives as a narrow correlated band whose profile edge sits just inside the septum aperture. In the right-hand panel each proton is assigned the most specific crystal process it underwent (precedence: volume capture $>$ channeling $>$ volume reflection $>$ amorphous, since a captured proton also carries channeling records and a channeled one volume-reflection records; untouched protons never met a crystal), so the classes are exclusive and sum to the extracted beam. Coordinates and interaction labels are recorded in the same tracking run.}
\label{fig:extbeam}
\end{figure*}

The match searches jointly the eight replay parameters not fixed by the as-built hardware: the normalised horizontal emittance $n_\mathrm{ex}$ of the initial distribution, which sets the separatrix structure and absorbs uncertainty in the extraction tune and closed orbit; the effective ZS interception width; and the three transverse and three angular per-crystal offsets of the array, which carry the designed stagger of Section~\ref{sec:concepts:mvra} together with whatever assembly and modelling residuals accompany it. The momentum distribution is a narrow Gaussian, $\sigma = 4.5\times 10^{-5}$, offset from the reference momentum: in the constant-optics slow extraction of the SPS the beam is swept onto the resonance by scaling all magnets together, so the extracted particles carry a common momentum offset~\cite{Kain_2019_PRAB}. Its centre, $+1.31\times 10^{-4}$, is set by requiring the model to extract the same fraction of the circulating beam as the measurement, and both are frozen before the search begins.

\subsection{Matching method and uncertainty budget}
\label{sec:expdata:method}

Each candidate is scored by a \emph{dip-region} metric, the mean absolute deviation between simulated and measured scans from the measured FWHM down to the measured minimum, which judges the shadowing well itself. The curves are \emph{not} dip-aligned: a bounded lateral offset is the single alignment degree of freedom and the reported deviation is its minimum over that offset, because aligning on the two minima would discard the information in where the dip sits and is fragile when the minimum lands on a noisy grid point. Every candidate is evaluated by a full position scan followed by an angle scan at the jaw position the position scan selects, at three particle-draw seeds, and the objective is the \emph{larger} of the two deviations, so that agreement on one scan cannot be bought at the expense of the other.

Because the two scans respond to different combinations of the eight parameters, the search was run as a global optimisation, Optuna~\cite{Akiba_2019_Optuna} driving a tree-structured Parzen estimator~\cite{Bergstra_2011_TPE} over the bounded ranges of Table~\ref{tab:searchspace}: $159$ trials ran to completion on the CERN batch farm, $151$ of them scored on both scans at three replay seeds, about $3\,400$ core-hours in all. Appendix~\ref{app:globalmatch} gives the optimiser, the search space and the convergence, and why the three defining choices (a global search, a free lateral offset, scoring on the worse scan) carry the result. The best three-seed configurations were re-scored on three seeds the optimiser had never used: the score of the configuration that led on the original seeds became $2.3$ times worse on the fresh ones, while the score of the adopted configuration changed by only $5\%$.

\subsubsection{What limits the comparison}

A simulated loss reduction without an uncertainty is not comparable to a measurement, and in a chain with this many stochastic stages the dominant term is only found by measuring each. Four terms were measured on the same observable, the loss at the alignment optimum normalised to the crystal-out loss (Table~\ref{tab:uncertainty}); every number in this subsection is an absolute standard deviation of that ratio, not a fraction of it. The largest is the capture of the initial distribution: independently seeded captures of the same beam at the same working point scatter by $0.041$, above the replay-seed term of every node and comparable to the $0.06$ separation between the low- and high-emittance groups of Section~\ref{sec:expdata:emitzs}. The map remains usable because its emittance ladder shares one capture seed, which leaves a row-to-row scatter of $0.018$. Appendix~\ref{app:budget} gives how each term was measured.

\begin{table}[h]
\squeezetable
\centering
\caption{Contributions to the uncertainty on the simulated loss at the alignment optimum, all measured on that quantity except the last capture row, which is at the fixed operating point (Appendix~\ref{app:budget}). The last row is the range spanned by the emittance/width map of Section~\ref{sec:expdata:emitzs}, from $0.169$ to $0.302$ across its $66$ nodes, i.e. the signal those terms have to be judged against.}
\label{tab:uncertainty}
\begin{tabular}{lr}
\hline
Contribution & Size \\
\hline
Replay seed, within one run & $0.005$--$0.03$ \\
Run to run, identical configuration & $0.008$ (max $0.028$) \\
Emittance row to row & $0.018$ \\
Capture, scan minimum ($n=4$) & $0.041$ \\
Capture, operating point ($n=356$) & $0.029$ \\
\hline
Range spanned by the map & $0.133$ \\
\hline
\end{tabular}
\end{table}

For Section~\ref{sec:expdata:emitzs} this means that a row mean is good to about $0.018$, while a single point-to-point difference smaller than about $0.03$ is not physics. Three further estimator choices were settled by measuring the alternatives, each time against the intuitive choice (Appendix~\ref{app:estimators}).

\subsection{Sensitivity to emittance and septum width}
\label{sec:expdata:emitzs}

\begin{figure}
\includegraphics[width=\columnwidth]{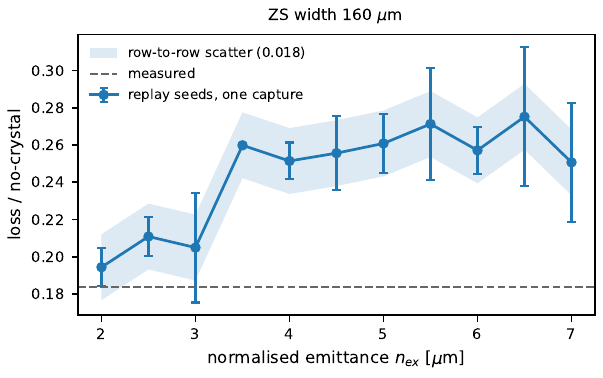}
\caption{Simulated loss at the alignment optimum, normalised to the crystal-out loss, against normalised emittance at the matched septum width of $160~\mu$m, from the $66$-point emittance / septum-width map (three seeds, six turns), with the measured value dashed. Error bars are the spread over replay seeds and the shaded band is the row-to-row scatter, which is the correct bar for comparing one emittance with another. The reduction keeps deepening below $3~\mu$m, while the agreement with the measurement is best \emph{at} $3~\mu$m and degrades on both sides of it (see the text).}
\label{fig:emit_zs_cut}
\end{figure}

Two of the eight matched parameters carry a physical interpretation beyond the fit: the normalised emittance, which sets the angular spread the crystals see, and the effective septum width. To separate their roles, and to test whether the chosen emittance is a real optimum or an artefact of the range the optimiser was given, we mapped the two on a grid with everything else frozen at the matched configuration. The grid runs over emittance from $2$ to $7~\mu$m in steps of $0.5~\mu$m against ZS width from $100$ to $400~\mu$m, $66$ points, each a full position scan followed by an angle scan at the jaw the position scan selects, at three seeds. The cut through that map at the matched septum width is Figure~\ref{fig:emit_zs_cut}.

The first result is negative: the modelled septum width does essentially nothing to the depth of the dip. Averaged over emittance, the loss at the six widths spans only $0.011$, less than the reproducibility of the pipeline, so there is no emittance/width degeneracy to trade against, and the width is fixed by the \emph{shape} of the scans. The emittance does set the depth, but the map's own noise allows only a comparison of groups. The rows from $2$ to $3~\mu$m shadow to $0.20$ of the crystal-out level and those from $3.5$ to $7~\mu$m to $0.26$, a separation three times the row-to-row scatter, whereas no single step between adjacent rows stands out from that scatter. We therefore report a low-emittance group that shadows more deeply than a high-emittance group and do not claim a threshold.

Because the depth of a volume-reflection dip depends steeply on the jaw position, we checked that the separation survives the choice of jaw at each node. The two independent conventions of Appendix~\ref{app:estimators} reproduce it point by point; only the third, which takes the jaw from the shallow amorphous position scan, is too noisy to resolve it. The grouped separation therefore does not depend on the jaw convention; it was computed on the common capture of the map and has not been repeated on a second capture seed.

The second result answers the question the map was built for, twice and in opposite directions. The optimisation had been given a lower bound of $3~\mu$m and the best configurations accumulated against it. Opening the range to $2~\mu$m shows that on \emph{loss} it is a wall: the reduction goes on deepening below $3~\mu$m, the mean of the lowest row, $0.19$, comes within the noise of the measured $0.184$ and individual points fall below it, so a low enough emittance lets the model out-shadow the machine. On \emph{agreement} the same grid puts an interior minimum at $3~\mu$m: the best dip-region deviation of the row is $0.038$, against $0.046$ half a micron below and $0.082$ half a micron above, so the matched value is not a boundary artefact. Each of these is a single three-seed value without an error bar, so the minimum is indicative only. The matched emittance is therefore where the model most resembles the machine across the shape of both scans; a model tuned to the measured depth alone would be driven to about $2~\mu$m and disagree with the scans everywhere else.

One caveat: the extraction front shrinks with the beam, so the low-emittance rows are the noisiest on the grid, with less than half the crystal-out counts of the high-emittance rows, which is why the number quoted is the row mean.

\subsection{Final matched configuration}
\label{sec:expdata:match}

\begin{figure*}
\centering
\includegraphics[width=0.92\textwidth]{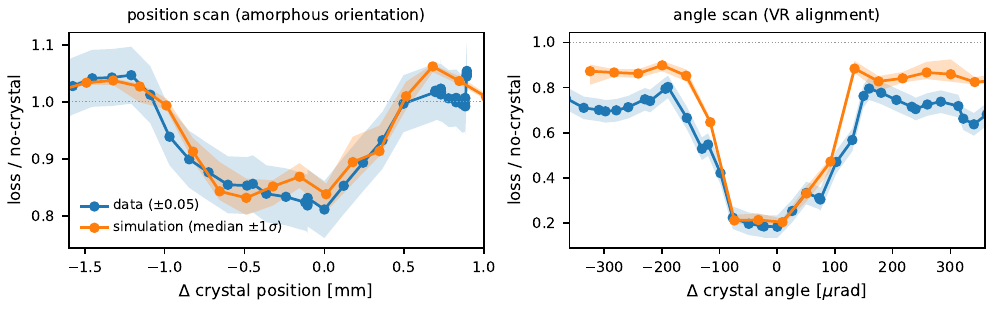}
\caption{Measured (blue, line with markers; the measured points carry no error bar, and the $\pm 0.05$ band is the position-overlap tolerance of the matching metric) and matched-simulation (orange, median over the six replay seeds of Table~\ref{tab:bestmatch} with the $16$th--$84$th percentile band) loss for the jaw (position, left) and angle (right) scans of the 3-crystal MVRA, both normalised to the same crystal-out loss so that the distance below unity is the shadowing itself. The curves are \emph{not} dip-aligned: the simulation is drawn at the lateral offset the matching metric selected, $+2.320$~mm on position and $+176~\mu$rad on angle, the simulation-to-machine calibration of the jaw and goniometer scales. \emph{Left:} the position scan at the amorphous machine-reference orientation, simulation $0.831$ against measured $0.811$. \emph{Right:} the angle scan, FWHM $205$ against $211~\mu$rad and minimum $0.204 \pm 0.041$ (capture; the mean of four independent captures is $0.26$) against $0.184$; off the dip the simulation sits at $0.88$ against $0.77$ (see the text). Over the dip region the two agree to $0.042$ on this normalisation and to $0.027$ on the AM baseline of the matching metric, the figure in Table~\ref{tab:bestmatch}.}
\label{fig:match_scans}
\end{figure*}

The final matched configuration is reported in Table~\ref{tab:bestmatch}: an initial distribution at $n_\mathrm{ex} = 3.0~\mu$m, an effective ZS wire-band width of $160~\mu$m, per-crystal transverse offsets of $[-35, -75, +15]~\mu$m and per-crystal angular offsets of $[-2.5, -22.5, +40]~\mu$rad. The overlay against the measured scans is shown in Figure~\ref{fig:match_scans}, both normalised to the crystals-retracted loss of Table~\ref{tab:denominators}.

\begin{table}[h]
\squeezetable
\centering
\caption{Final matched simulation configuration, from the global optimisation described in the text. The upper block lists the eight jointly matched parameters and the lower block the quantities held fixed at their as-built or nominal values. The jaw (position) scan is taken with the array at its amorphous machine-reference orientation. Jaw positions are in the simulation frame, which runs opposite to the machine's: a machine position is $-x_\mathrm{sim} - 2.320$~mm, the second term being the fitted calibration of Figure~\ref{fig:match_scans}. The two deviations are the mean $|\mathrm{simulation}-\mathrm{machine}|$ over the dip region, averaged over six replay seeds: the three of the search and the three of the fresh-seed re-scoring.}
\label{tab:bestmatch}
\begin{tabular}{lr}
\hline
Matched parameter & Value \\
\hline
Normalised emittance $n_\mathrm{ex}$ & $3.0~\mu$m \\
ZS effective wire-band width & $160~\mu$m \\
Per-crystal transverse offsets & $[-35,\,-75,\,+15]~\mu$m \\
Per-crystal angular offsets & $[-2.5,\,-22.5,\,+40]~\mu$rad \\
\hline
Fixed parameter & Value \\
\hline
Strip width / length / bend & $1.25$ / $2$~mm / $160~\mu$rad \\
Momentum spread $\delta p/p$ & Gaussian, $\sigma = 4.5\times 10^{-5}$ \\
 & centred at $+1.31\times 10^{-4}$ \\
Crystal jaw position, angle scan & $-39.625$~mm \\
\hline
Dip-region deviation & Value \\
\hline
Position scan & $0.019$ \\
Angle scan & $0.027$ \\
\hline
\end{tabular}
\end{table}

\begin{figure*}
\centering
\includegraphics[width=0.92\textwidth]{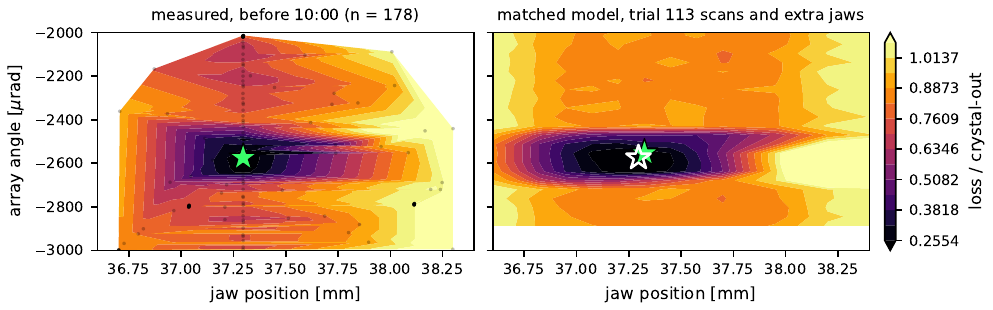}
\caption{Held-out test of the matched model against the two-dimensional scan. \emph{Left:} the measured first-period map of Figure~\ref{fig:sobol_contour} ($178$ cycles), normalised to the crystal-out loss. \emph{Right:} the loss predicted by the adopted configuration of Table~\ref{tab:bestmatch} (thirteen angle scans, three replay seeds averaged, six turns, the same normalisation), placed on the measured axes at the calibration offsets of the one-dimensional scans. Stars mark the loss optimum of each map, the open star on the right being the measured one; the panel titles refer to the 10:00 local-time split of Figure~\ref{fig:sobol_contour}. The depth carries the capture systematic of Table~\ref{tab:uncertainty}.}
\label{fig:sobol_heldout}
\end{figure*}

An angle scan divided by its own AM baseline is pinned to unity off the dip by construction, so measurement and simulation agree there whatever they do; normalised to the crystal-out loss, the off-dip level instead measures how much the array shadows before it is aligned. The position scan brackets the retracted condition directly. For the angle scan, taken at a fixed jaw with the crystal in, the reference is the retracted-crystal track in the simulation and, in the machine, the retracted end of the position scan taken twenty minutes earlier.

Against that reference the matched simulation reproduces both dips, and the two kinds of agreement read differently. The \emph{shape} is the robust part, because the capture term of Table~\ref{tab:uncertainty} moves a dip up or down without moving its width or position. Both widths are reproduced to $3\%$. Over the dip region the mean absolute deviation between the profiles is $0.019$ (position) and $0.027$ (angle) in the normalised loss of the metric, and $0.042$ for the angle scan normalised to the crystal-out loss; over every measured point the simulated grid covers, it is $0.026$ and $0.044$. These are absolute differences of the normalised loss, not relative errors. The \emph{depth} carries that systematic. On the capture the match was run on, the angle dip reaches $0.20 \pm 0.04$ of the crystals-retracted loss against the measured $0.184$. That capture is the lowest of the four independent ones of Appendix~\ref{app:budget}, whose mean is $0.26 \pm 0.02$ (standard error), so on average the model under-shadows at the minimum by about $0.08$, in the same direction as the off-dip residual below; the comparison of depths is limited by the initial distribution. At the operating point the $356$-capture ensemble gives $0.26 \pm 0.03$, against $0.30$ to $0.31$ in routine operation on the same reference (Sections~\ref{sec:feedback:deployment} and~\ref{sec:setup:obs}).

The angle scan carries a second, sharper feature that measurement and matched scan both sample too coarsely to resolve. Re-scanning the simulated angle in $5~\mu$rad steps at six seeds shows that the broad dip is not flat at its floor: a structure about $15~\mu$rad wide sits inside it and reaches $0.177$ of the no-crystal loss, some $0.05$ below the plateau that surrounds it over at least $45~\mu$rad. The measured scan, stepped at $20~\mu$rad, bottoms at $0.184$ on the same reference, so with its uncertainty the comparison is $0.177 \pm 0.041$ (capture) against $0.184$. That $15~\mu$rad optimum is a prediction of the model, which the measured scan cannot resolve; it is no wider than the single-crystal dip that Section~\ref{sec:concepts:nonlocal} identifies as too narrow to hold open-loop, so it could not be the setpoint in any case. The array is operated on the surrounding plateau, several times wider (the reduction exceeds $4\times$ over $105~\mu$rad, Section~\ref{sec:expdata:meas}), which is the wide-acceptance argument of Section~\ref{sec:concepts:mvra} in quantitative form, and the median of the operational loss distribution of Section~\ref{sec:feedback:deployment} sits on the plateau. The small depth disagreement on the angle scan changes sign between the two normalisations, because the simulation's higher off-dip level inflates its own AM baseline.

Two residuals remain. The angle scan carries one large local excursion, $0.20$, on the steep rising flank of the dip, where the simulation reaches $0.94$ against a measured $0.74$: on a flank that steep a small horizontal misplacement produces a large vertical difference, so the excursion measures the alignment sensitivity of the flank and does not constrain depth or width. The second is the off-dip level, $0.88$ of the no-crystal loss in the model against $0.77$ measured: at the angle-scan jaw the array in its amorphous orientation already removes about a fifth of the septum loss in the machine and about half that in the model. This is the same under-shadowing the amorphous-orientation jaw scan shows, and the clearest remaining target for the model.

The eight parameters and the two calibration offsets were fitted to $25$ scored points of the two one-dimensional scans, so the agreement in Figure~\ref{fig:match_scans}, widths included, is a fit: the metric runs from the measured half maximum to the minimum, so the width of each dip enters the objective. The independent test is the two-dimensional map.

The two-dimensional map of Figure~\ref{fig:sobol_contour} played no part in the fit and is a test the one-dimensional match cannot supply. Figure~\ref{fig:sobol_heldout} sets the measured first-period map beside the prediction of the adopted configuration: thirteen angle scans at jaw positions $0.125$ to $0.25$~mm apart, drawn on the measured axes at the calibration offsets already fixed on the one-dimensional scans, $+2.320$~mm and $+176~\mu$rad, with no parameter moved. Nine of the scans were computed during the search as trial 113's bracketed jaw sweep, of which only the scan at the selected jaw entered the objective; the other four were added afterwards under the same protocol. The one-dimensional scans that fixed the parameters were taken between 06:30 and 07:10 that morning, in the same SFTPRO-only supercycle as the first period. The new information is two-dimensional: the location of the optimum, the spatial width of the basin, its tilt and its flanks; the angular width at the fitted jaw repeats the one-dimensional match. The grid spans $1.9$~mm in position and covers $159$ of the $178$ first-period cycles, the remainder lying below its angular edge.

The predicted minimum, at $37.33$~mm and $-2554~\mu$rad, lies a quarter of a jaw step and half an angle step of the grid from the measured one at $37.30$~mm and $-2574~\mu$rad, and the three replay seeds place it within one jaw step of each other. The angular width of the basin is $200$ to $207~\mu$rad across the seeds against $209~\mu$rad measured the same way. The spatial width through the minimum, taken at the crystal-out level, is $0.95$ to $0.97$~mm across the seeds; the measured map is too sparse near its optimum to give the same number, and the $1.276$~mm of the position scan belongs to the shallow amorphous-orientation dip. Both fitted quadratics have an axis ratio of $1.3$ and a weak tilt of opposite sign ($+27$ to $+55~\mu$rad/mm predicted against $-12$ measured over a $\pm 0.6$~mm by $\pm 150~\mu$rad window). Over the $17$ covered cycles inside the basin (loss below $0.5$ of the crystal-out level) the prediction sits $0.054$ above the measurement on average, with a mean absolute deviation of $0.065$; over all $159$ covered cycles it is $0.13$, dominated by the amorphous wings where the model already sits high. The prediction fails on the high-position flank, which closes about $0.1$~mm early in the model, at $37.73$~mm against about $37.85$~mm measured: the two cycles at $37.74$~mm read $0.33$ and $0.41$ against a predicted $0.54$ and $0.62$, and without them the basin deviation is $0.046$. With that exception, the matched model predicts the two-dimensional structure of a measurement its objective never saw.

The matched per-crystal angular offsets, $[-2.5, -22.5, +40]~\mu$rad, admit two independent checks against the array's designed stagger (Section~\ref{sec:concepts:mvra}).

The first is direct. Autocollimator metrology on the original four-strip assembly, performed by the crystal-production programme before installation, measured the relative tilts between successive strips as $+300$, $+14$ and $+60~\mu$rad~\cite{Aberle_2026_Channeling}: three strips within $74~\mu$rad of one another and a fourth $300~\mu$rad away, outside any orientation at which it could contribute coherently. A test-beam scan reproduced that structure, the outlying strip was removed, and the three-strip array was revalidated in the test beam before installation. The fitted offsets are relative: their common part, $+5~\mu$rad, cannot be separated from the $+176~\mu$rad angular calibration of the scan (Appendix~\ref{app:globalmatch}), and likewise the common part of the transverse offsets from the jaw calibration, so only the differences between strips carry information. The fitted offsets span $62.5~\mu$rad against the measured $74~\mu$rad, and their largest step between successive strips, $62.5~\mu$rad, sits close to the largest measured step of $60~\mu$rad; the smaller step differs more, $20$ against $14~\mu$rad, and its sign depends on a mapping between the simulation's strip indices and the metrology's that we have not fixed. Read at the level of magnitude and ordering, a fit given a free $\pm 50~\mu$rad on each strip and no knowledge of the metrology returned a stagger of the size and shape that was built. We nevertheless report the offsets as \emph{effective} parameters, because without miscut and torsion metrology on these strips the fit cannot separate the stagger from model error absorbed into the nearest degrees of freedom.

The second is the width of the dip, a model-dependent consistency check rather than an independent measurement of the stagger. A proton reaches each strip after the kicks of the strips before it, so the acceptance of the array is not simply the offsets added to the $160~\mu$rad swept by each bend; we therefore counted, in the matched model at the operating jaw, the strips in which each traversing proton is volume-reflected on its first passage through the array, at every array angle of a $10~\mu$rad scan. The fraction reflected in \emph{at least one} strip exceeds one half over $213~\mu$rad, in \emph{at least two} over $155~\mu$rad, and in \emph{all three} over $97~\mu$rad, reaching at most $84\%$; interval arithmetic on the offsets alone gives $222.5$, $160$ and $97.5~\mu$rad, so the cumulative kicks shift the windows by less than a scan step. The simulated dip of the same scan has a FWHM of $201~\mu$rad and the measured one $211~\mu$rad, both between the at-least-two and at-least-one windows: the reduction persists while any strip is still reflecting and fades as the last one leaves. An unstaggered array would give a single window of about $160~\mu$rad on all three counts, some $50~\mu$rad narrower than the measurement. Widening the working range is what the stagger is for, and the measured width is consistent with it.

\section{On-line Control with Extremum Seeking}
\label{sec:feedback}

\subsection{Operational drift problem}
\label{sec:feedback:problem}

The optimum of Section~\ref{sec:expdata} is an open-loop position-and-angle setpoint found during a dedicated MD. In production it does not stay put: the closed orbit, the emittance and the mechanical position of the ZS girder drift on a timescale of hours to days, and each moves the loss optimum, as Figure~\ref{fig:sobol_contour} caught happening within an hour when the supercycle composition changed. The MVRA therefore needs an on-line controller that holds the setpoint at the local loss optimum and re-acquires it when drift moves it away.

\subsection{Choice of an extremum-seeking controller}
\label{sec:feedback:choice}

Extremum seeking (ES) is a model-independent feedback method with a long history~\cite{Scheinker_2024_survey}. Its bounded-update form~\cite{Scheinker_2014_bounded} was brought to particle accelerators in~\cite{Scheinker_2013_PRSTAB} and has been used at CERN to tune the AWAKE electron line~\cite{Scheinker_2020_AWAKE}. An ES controller was chosen here because the measured loss response around the optimum is approximately quadratic (Figures~\ref{fig:sobol_contour} and~\ref{fig:match_scans}), so the scalar loss alone carries enough gradient information to track the optimum without a model of the surface. The wide, single, near-quadratic basin of the array does not need the surrogate-guided and model-based controllers used for the single-crystal devices.

\subsection{Algorithm and SPS adaptation}
\label{sec:feedback:algo}

The controller is the outer loop of a cascade. It is built around the \texttt{ExtremumSeeker} class of the CERN ML extremum-seeking library~\cite{cernml_ES_lib}, the bounded-update extremum seeking of~\cite{Scheinker_2014_bounded,Scheinker_2021}, in which the measured loss enters the \emph{phase} of the dither rather than a gradient estimate. With $J_k$ the scalar loss Eq.~\eqref{eq:tot_loss} at iteration $k$, each coordinate $i$ of the (position, angle) setpoint is moved by
\begin{equation}
x_{i,k+1} = x_{i,k} + a_k\, \omega_i \Delta t\, d_i \cos\!\left(\omega_i \Delta t\, k + g J_k\right),
\label{eq:es_update}
\end{equation}
with dither size $d_i$, gain $g$, amplitude $a_k$ (Eq.~\eqref{eq:es_amplitude} below), distinct frequencies $\omega = 1$ and $1.75$ for the two coordinates, and $\Delta t = 2\pi/(5 \times 1.75)$, i.e.\ five samples per period of the faster dither. On average this drifts the setpoint down the loss gradient while dithering around it. The update runs in a scaled $[-1, 1]^2$ action space that is mapped to the physical coordinates at the boundary. The controller takes at most one step per SPS extraction cycle, and only when the inner proportional-integral-derivative (PID) loop that drives the goniometer reports that the previous setpoint was reached and held through the extraction; otherwise the cycle is skipped and the seeker holds. Table~\ref{tab:es_settings} lists the settings retained for operation; at unit amplitude the dither is $60~\mu$m and $18~\mu$rad, a small fraction of either dip width (Section~\ref{sec:expdata:meas}). Parts of the controller software were developed with an AI coding assistant, as stated in the Acknowledgments.

\begin{table}[h]
\squeezetable
\centering
\caption{Settings of the extremum-seeking controller retained for operation, in effect from 31 July 2026 to the end of the run on 31 August; the routine-operation record of Section~\ref{sec:feedback:deployment} predates them and was taken under the wider-dither schedules named in Section~\ref{sec:feedback:algo}. Scaled quantities refer to the $[-1, 1]^2$ action space of the search box; the physical dither is the amplitude times the dither size times the half-width of the box.}
\label{tab:es_settings}
\begin{tabular}{@{}p{0.58\columnwidth}@{\hspace{6pt}}l@{}}
\hline
Parameter & Value \\
\hline
Gain & $0.4$ \\
Dither, position / angle (scaled) & $0.2$ / $0.06$ \\
Samples per dither period & $5$ \\
Cost target $J_\mathrm{t}$ & $0.22$ \\
Amplitude $a_\mathrm{min}$ / $a_\mathrm{max}$ & $0$ / $0.4$ \\
Sigmoid midpoint $m$ / sensitivity $s$ & $0.10$ / $80$ \\
Cost smoothing window $N$ & $5$ iterations \\
Search box, position & $37.2$ to $37.8$~mm \\
Search box, angle & $-2800$ to $-2200~\mu$rad \\
Deadband, position / angle & $10~\mu$m / $10~\mu$rad \\
Advance only after the PID settles & yes \\
Watchdog: amplitude fraction & $1.05$ \\
Watchdog: iterations / improvement & $150$ / $0.05$ \\
Rail guard: clamped iterations & $10$ \\
Automatic restarts before latching & $2$ \\
Model seed & off \\
\hline
\end{tabular}
\end{table}

The loss-dependent amplitude adjustment is a sigmoid schedule on an exponentially smoothed cost estimate. With $J_k$ as in Eq.~\eqref{eq:es_update}, the smoothed cost and the dither amplitude are
\begin{equation}
\begin{aligned}
\tilde{J}_k &= \tilde{J}_{k-1} + \frac{J_k - \tilde{J}_{k-1}}{N}, \\
a_k &= a_\mathrm{min} + \frac{a_\mathrm{max} - a_\mathrm{min}}{1 + \exp\!\left[-s\left(\left|\tilde{J}_k - J_\mathrm{t}\right| - m\right)\right]},
\end{aligned}
\label{eq:es_amplitude}
\end{equation}
with the values of Table~\ref{tab:es_settings} in scaled action units. The routine-operation record of Section~\ref{sec:feedback:deployment} was accumulated under earlier, wider-dither schedules of the same controller; the longest-running of them (8 to 25 July: dither $0.33$ and $0.088$ of the half-widths, $a_\mathrm{max} = 1.5$, $s = 50$, $J_\mathrm{t} = 0.20$) kept a dither of $40$ to $60~\mu$m alive on the plateau, which the retained settings remove. No per-cycle statistic is reported for the weeks they were in use, so they are characterised by the following calculation at the loss levels of the earlier record.

The amplitude is smallest at the target and grows on either side of it. At the record's median loss of $0.278$ and mean of $0.291$ the smoothed cost sits $0.06$ to $0.07$ above the target and the amplitude is $3$ to $9\%$ of its maximum. That is a dither of $1$ to $2~\mu$m in position and below $1~\mu$rad in angle, well inside the $10~\mu$m / $10~\mu$rad hardware quantisation of Section~\ref{sec:setup:diagnostics}. Under the retained settings the dither alone therefore cannot cross the deadband, and the schedule issues no exploratory moves on the plateau. The amplitude reaches half its maximum at a smoothed loss of $0.32$, where the position dither crosses the deadband, and $95\%$ of it at $0.36$; at full amplitude the dither is $24~\mu$m and $7~\mu$rad. The dither re-grows within a few iterations once a drift has raised the smoothed loss to about $0.32$, that is $0.10$ above the cost target and $0.04$ above the median of the record, which is the detection threshold noted in Section~\ref{sec:activation:limits}. The smoothing over $N$ iterations keeps single-iteration noise from triggering a spurious wake-up. The hardware quantisation is enforced explicitly: a proposed move below the deadband on both axes is not issued, the controller re-observes at the same setpoint, and the seeker keeps advancing its internal state, which is never re-anchored to the measured goniometer position, without burning a hardware move. A move is issued once that internal state lies more than one quantum from the PID setpoint on either axis, so a correction can accumulate on an axis whose instantaneous dither alone stays below its quantum.

Two guards protect the loop, added after a runaway on 3 July 2026 in which beam returning after a stop with a loss floor of $0.43$ saturated the amplitude and drove the crystal from rail to rail for $5.8$~h. A watchdog trips after $150$ consecutive iterations with the amplitude saturated and no improvement of the smoothed cost by $0.05$, and a rail guard after $10$ consecutive proposals clipped to the search box; a trip triggers up to two automatic clean restarts from the best point seen, after which the loop latches and waits for an operator. Both guards went into operation on 8 July 2026, so the first two weeks of the routine record ran without them; until the retained settings they used a saturation threshold of $0.9$ of the maximum amplitude over $40$ iterations. With the threshold raised to $1.05$ of the maximum amplitude in the retained settings the watchdog cannot fire, and the rail guard is the active protection. The surrogate-model seed of the deployment is disabled, and the loop starts from the operator's setpoint.

In closed loop the per-cycle loss falls from $\approx 0.4$ to $\approx 0.27$ as the controller walks the setpoint into the optimum, the dither large during acquisition and decaying once the loss settles onto its plateau.

\subsection{Operational deployment and sustained performance}
\label{sec:feedback:deployment}

\begin{figure}
\includegraphics[width=\columnwidth]{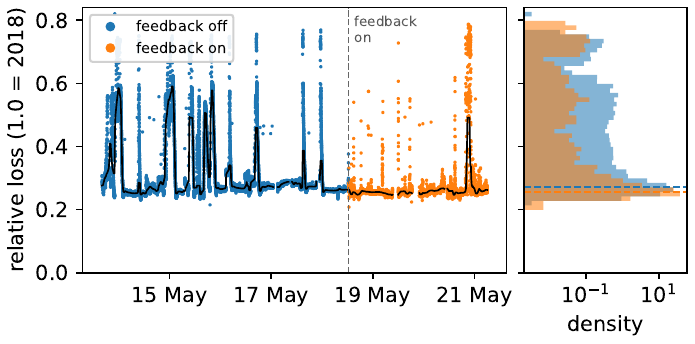}
\caption{The May 2026 operational test: per-cycle intensity-normalised loss Eq.~\eqref{eq:tot_loss} versus time (left; feedback off in blue, on in orange, rolling median in black, the dashed line marking the feedback-on transition) and the corresponding loss distributions (right, log density). Feedback narrows the distribution (standard deviation $0.098 \to 0.054$; $22\,026$ and $10\,366$ cycles) while holding the median near $0.26$.}
\label{fig:fb_hist}
\end{figure}

The controller was deployed on the production slow-extraction beam, first in shadow mode and then with feedback to the hardware enabled. Figure~\ref{fig:fb_hist} shows the May operational test ($22\,026$ cycles with feedback off and $10\,366$ with it on, SFTPRO1 cycles above $10^{13}$ protons with the array inserted, split at the feedback-on transition of 18 May 2026). The feedback does not deepen the loss minimum, the median moving only from $0.27$ to $0.26$, but it narrows the distribution sharply: the standard deviation of the per-cycle loss falls from $0.098$ to $0.054$, a $\sim 45\%$ reduction, as the controller suppresses the recurrent upward excursions of the feedback-off phase, each a drift of the operating point off the optimum. The two phases are consecutive, not interleaved, so part of the narrowing could come from a quieter machine.

\begin{figure}
\includegraphics[width=\columnwidth]{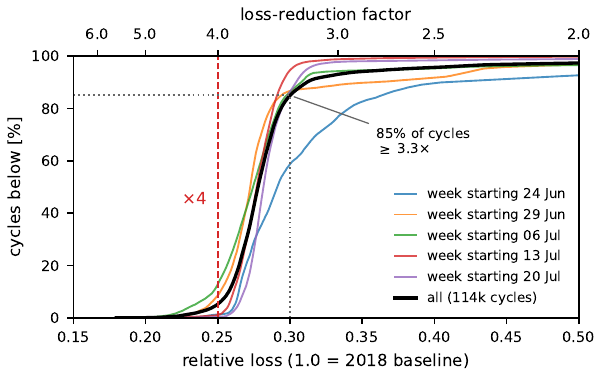}
\caption{Cumulative distribution of the per-cycle intensity-normalised loss Eq.~\eqref{eq:tot_loss} over the routine-operation period of the MVRA (SFTPRO1 cycles with the array inserted, from its deployment in operation on 24 June 2026 to 24 July, $1.14\times 10^5$ cycles), overall (black) and by week. Loss and reduction factor are on the 2018 pre-shadowing baseline of Eq.~\eqref{eq:tot_loss}; the upper axis gives the equivalent loss-reduction factor, the dashed line marks the factor-of-four target and the dotted guide the $85\%$ level. The median is $0.278$ ($3.6\times$) and $85\%$ of cycles achieve $3.3\times$ or better; the weekly curves nearly coincide.}
\label{fig:fb_ts}
\end{figure}

The MVRA and its controller entered routine operation on 24 June 2026 and ran in closed loop on the production beam until the loop was switched off on 31 August. Figure~\ref{fig:fb_ts} gives the cumulative distribution of the per-cycle loss over the $1.14\times 10^{5}$ production cycles delivered with the array inserted from 24 June to 24 July, under the earlier dither schedules of Section~\ref{sec:feedback:algo}. The sample is selected only on the array being inserted and on the extracted intensity, so it keeps every cycle in the window, including the 3 July runaway and any period with the feedback disabled. The median relative loss on the 2018 baseline of Eq.~\eqref{eq:tot_loss} (Table~\ref{tab:denominators}) is $0.278$, a sustained reduction of $3.6\times$; $85\%$ of cycles lie below $0.30$, a reduction of $3.3\times$ or better, and $94\%$ below $0.35$. The distribution has a tail towards high loss, so its mean, $0.291$ ($3.44\times$), sits above the median; activation integrates the delivered loss, so the mean is the figure for an activation projection (weighted by extracted intensity it is also $0.291$, over $1.5\times 10^{18}$ extracted protons, with a day-to-day bootstrap spread of $0.006$), while the median describes where the controller holds the machine on a typical cycle. On the crystals-retracted reference of the design the two numbers are $0.30$ and $0.31$ (Section~\ref{sec:setup:obs}), reductions of $3.3\times$ and $3.2\times$; both sit below the factor-of-four target, a point returned to in Section~\ref{sec:activation:limits}. The weekly medians span only $0.272$ to $0.291$ over five weekly bins, the first and last partial, the first (commissioning) bin the broadest and the later weeks indistinguishable. The controller held this operating point through continuous physics production with one manual intervention, the restart after the 3 July runaway of Section~\ref{sec:feedback:algo}.

\section{Loss Budget and Limitations}
\label{sec:activation}

\subsection{Loss budget across the multi-year programme}
\label{sec:activation:budget}

To compare the shadowing configurations on a common footing we use a monitor-based, collateral-aware figure of merit against a single no-crystal reference, the crystals-retracted reference of Table~\ref{tab:denominators}. The local (TECS) and non-local (TECA-single, MVRA) crystals sit in different straight sections and are operated alternately, so the crystal-out cycles of a non-local crystal's own window still have the local crystal in, and referencing to them measures the non-local crystal against the local one. Each configuration is therefore referenced to the true no-crystal state, obtained by pooling the brief both-retracted transitions of the MD programme across the 2024 and 2025 windows, $55$ cycles in all. Against this reference we define, per cycle, the normalised ZS/TCE loss $L_\mathrm{zs}$ over the six LSS2 BLMs of Eq.~\eqref{eq:tot_loss} and the normalised collateral loss $D$ over the $103$ other loss monitors of sextants 2 and 4, covering LSS2, LSS4, the arcs between them and the TT20 transfer line and excluding only the TDC2 splitter cluster, whose losses are set by the extracted-beam optics and do not respond to the crystal. The collateral-charged figure of merit $M = (1 - L_\mathrm{zs})/D$ credits the septum protection and charges any loss the shadowing displaces elsewhere ($M = 0$ for no shadowing, $M = 0.75$ for an ideal fourfold ZS reduction at unchanged collateral). Table~\ref{tab:budget} reports the medians for each configuration, with the sample behind each row.

\begin{table}[h]
\centering
\caption{Collateral-aware loss budget across the SPS slow-extraction shadowing configurations, evaluated against a common no-crystal reference (both crystals retracted). $1-L_\mathrm{zs}$ is the reduction of the ZS/TCE (LSS2) loss (with the equivalent reduction factor); $D$ is the collateral loss elsewhere in the extraction sextants relative to the same reference ($D=1$ is unchanged collateral); $M = (1-L_\mathrm{zs})/D$ is the collateral-charged figure of merit. Medians over SFTPRO1 cycles above $10^{13}$ protons from the multi-year archive analysed in this work: TECS, 2024, $n = 722\,519$; TECA-single, 21 to 28 October 2025, $n = 29\,714$; MVRA, 13 to 24 May 2026, $n = 33\,067$; the reference, $55$ both-retracted cycles of 2024 and 2025 (for the choice of reference window see the text). $M$ is the median of the per-cycle ratio, so it does not reproduce exactly from the medians in the other two columns. The MVRA row is the May sample, selected with the crystal-in, TECS-out gate of this analysis and therefore not identical to the $22\,026 + 10\,366$ cycles of the feedback test of Section~\ref{sec:feedback:deployment}, which use a crystal-position cut and the feedback split; the June to July routine-operation median of Section~\ref{sec:feedback:deployment} is $0.30$ on this reference.}
\label{tab:budget}
\begin{tabular}{lccc}
\hline
Configuration & $1-L_\mathrm{zs}$ & $D$ & $M$ \\
\hline
No-crystal reference & $\equiv 0$ & 1.00 & 0.00 \\
TECS (local) & 0.20 ($1.2\times$) & 0.84 & 0.23 \\
TECA-single (non-local) & 0.31 ($1.5\times$) & 0.76 & 0.41 \\
MVRA (non-local array) & 0.72 ($3.6\times$) & 0.96 & 0.74 \\
\hline
\end{tabular}
\end{table}

\begin{figure}
\includegraphics[width=\columnwidth]{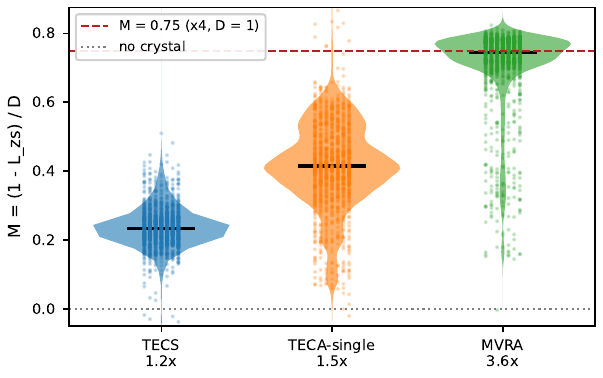}
\caption{Distribution of the collateral-aware figure of merit $M = (1-L_\mathrm{zs})/D$ across operational cycles for each shadowing configuration (collateral $D$ over the $103$ sextant-2 and sextant-4 monitors). Horizontal lines mark $M=0$ (no shadowing) and $M=0.75$ (an ideal fourfold ZS reduction at unchanged collateral). The ordering MVRA $\gg$ TECA-single $>$ TECS is clear, with the MVRA distribution concentrated near the ideal value.}
\label{fig:violin}
\end{figure}

The three configurations order MVRA $\gg$ TECA-single $>$ TECS (Figure~\ref{fig:violin}): the array reduces the ZS/TCE loss by a factor of $3.6$, to $0.28$ of the no-crystal level in this May sample (its closeness to the $0.278$ of routine operation, which is on the 2018 baseline, is a coincidence of the two references), against $1.5$ and $1.2$ for the non-local and local single crystals, and reaches $M = 0.74$ against the ideal-fourfold $0.75$. The single-crystal reductions are smaller than the $\sim 25\%$ (local) and $\sim 50\%$ (non-local) previously reported~\cite{Velotti_2024_IPAC}, for different reasons. For the local crystal it is the reference: $25\%$ against the 2018 level is $20\%$ against a no-crystal level that reads $0.93$ of it. For the non-local crystal, the $\sim 50\%$ was measured at the aligned optimum of a dedicated MD against the crystal-out cycles of its own window, that is, against the machine with the local crystal in, whereas the sample of Table~\ref{tab:budget} is a production week in October 2025 whose median sits at $0.69$ of the no-crystal level and whose lowest five per cent of cycles reach $0.60$. The two conditions differ, and this paper does not resolve the gap between them.

\begin{figure*}
\centering
\includegraphics[width=0.92\textwidth]{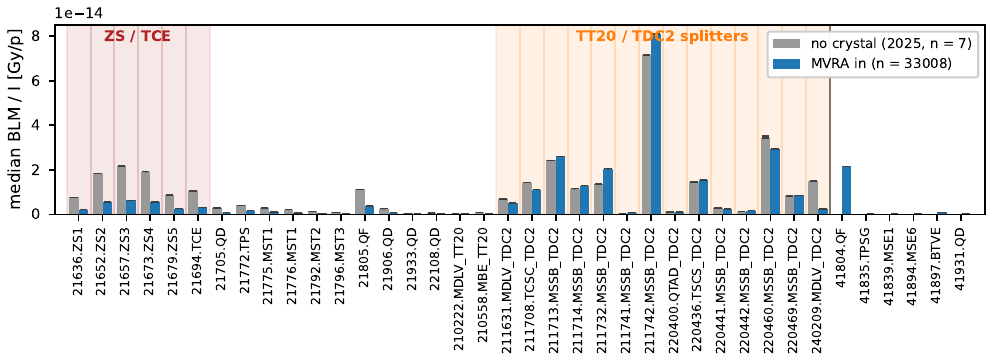}
\caption{Per-BLM intensity-normalised loss along the SPS extraction region, including the TT20 transfer line and TDC2 splitter cluster (shaded), with the MVRA inserted (blue, $n = 33\,008$ cycles, the MVRA cycles of the within-2026 control that fall in the $1.2$ to $1.6\times 10^{13}$ intensity band of the no-crystal reference) and with no crystal (grey, the seven both-retracted cycles of the 2025 window, intensity-matched; error bars are $\pm 1\sigma$). The ZS/TCE monitors (left shaded band) fall with the array in; the TT20/TDC2 splitter losses are essentially unchanged; the volume-reflected beam is swept onto an LSS4 monitor downstream of the array.}
\label{fig:blm_profile}
\end{figure*}

Figure~\ref{fig:blm_profile} shows where the MVRA puts the intercepted beam. Along the extraction region, TT20 and its TDC2 splitter cluster included, the ZS/TCE monitors drop sharply with the array in, the TT20/TDC2 losses are essentially unchanged, and a single LSS4 monitor a few metres downstream of the array picks up the volume-reflected beam swept out of the extraction path. The collateral factor $D$ is below one for the single crystals and $0.96$ for the array because $85\%$ of the collateral loss in the no-crystal reference sits in LSS2 downstream of the ZS, on the monitors of the thin-septum protection absorber (TPS), of the thin magnetic septa (MST) and of the neighbouring quadrupoles; that loss is fed by protons scattered off the septum wires and falls with the shadowing, to $0.88$, $0.66$ and $0.30$ of the reference for TECS, TECA-single and the MVRA. For the array this gain is cancelled by the new loss at LSS4, so its $D$ of about one is the net of two opposite changes: the septum is protected by relocating the intercepted beam within LSS4. The reference window matters at the $15\%$ level: the $48$ cycles of 2024 carry $25\%$ more collateral per proton than the seven of 2025, mostly in the sextant-2 arc, so against the 2025 cycles alone $D$ reads $0.99$, $0.89$ and $1.13$ and $M$ reads $0.17$, $0.32$ and $0.62$; the ordering and the size of the MVRA gain do not depend on the choice of reference.

Because this comparison spans years it inherits the recalibration caveat of Section~\ref{sec:setup:obs}. A control taken entirely within 2026 reaches the same conclusion free of it: comparing the TECS-only periods ($n = 18\,723$ SFTPRO1 cycles above $10^{13}$ protons) with the MVRA periods ($n = 33\,067$ cycles, 13 to 24 May 2026) in the same machine configuration and monitor calibration, the ZS/TCE loss with the MVRA drops to $0.34$ of the TECS-only level while the summed TT20/TDC2 loss ratio is $1.02$.

\subsection{Total loss reduction in simulation}
\label{sec:activation:simtotal}

\begin{figure}
\includegraphics[width=\columnwidth]{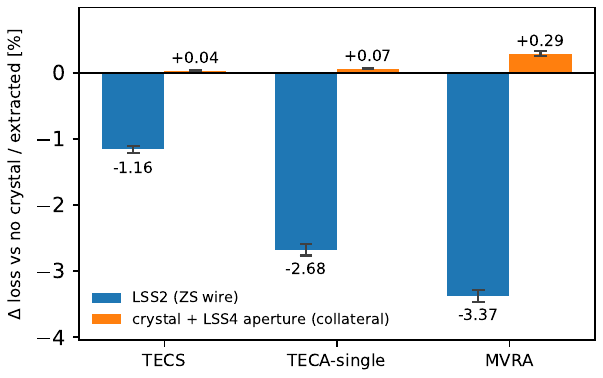}
\caption{Simulated decomposition of the total-loss change relative to the no-crystal baseline (Xsuite replay on the pre-match model, $400~\mu$m septum, as-operated extraction-bump optics; see the text; error bars are the spread over eight independent replays, each on its own $4\times 10^4$-particle draw from the captured initial distribution): change in loss per extracted proton, split into the LSS2 ZS-wire term (blue, the septum gain) and the crystal-inelastic plus LSS4-aperture collateral (orange), for the three configurations. The collateral cost is an order of magnitude smaller than the septum gain for every method, so the total loss falls for every method, least with TECS and most with the MVRA ($24$, $61$ and $72\%$).}
\label{fig:sim_split}
\end{figure}

The measured figure of merit is a sum of monitor responses over the extraction region, and equal summed response is not the same as equal numbers of lost protons or equal activation, so it bounds without closing the question of whether the \emph{total} extracted-beam loss falls. The Xsuite simulation, which follows every proton of the replay to its fate and pairs each configuration against a no-crystal baseline on the same seed, indicates that it does: all three configurations reduce the total loss, by $24\%$ for the local single crystal, $61\%$ for the non-local single crystal and $72\%$ for the array. Figure~\ref{fig:sim_split} shows why: the collateral each adds (crystal-inelastic plus LSS4 aperture) is an order of magnitude below the septum loss it removes. The MVRA generates the most collateral of the three, about four times that of a single crystal, and still gives the largest net reduction; its ordering against the single crystals reproduces the measured one of Section~\ref{sec:activation:budget}. The sizes do not all agree: the simulated $61\%$ for the non-local single crystal is above both the $\sim 50\%$ of the dedicated measurement and the $31\%$ of the production week. The comparison was made on the pre-match beam model with a $400~\mu$m septum and has not been recomputed on the matched model, so its absolute wire-hit fractions are not those of Section~\ref{sec:expdata:matching}. The septum is a perfect absorber in this model, so the protons scattered off the wires, which feed most of the reference collateral (Section~\ref{sec:activation:budget}), are not represented. What it establishes is the ordering and the separation between septum gain and collateral cost.

\subsection{Limitations}
\label{sec:activation:limits}

The limitations stated in the preceding sections are collected here with the hardware-related ones.

\paragraph{Performance against the factor-of-four target.} Routine operation under feedback holds a reduction of $3.4\times$ (mean) to $3.6\times$ (median) on the 2018 baseline against which the factor-of-four target was set, and $3.2\times$ to $3.3\times$ on the crystals-retracted reference of the design (Table~\ref{tab:denominators}, Section~\ref{sec:feedback:deployment}). Part of the gap is the choice of setpoint, since the array is operated on the plateau and not on the narrower minimum (Section~\ref{sec:expdata:match}), and part is the cost of holding a drifting optimum with a quantised goniometer; closing it needs finer controller resolution and the 4- or 5-crystal upgrade. The best point of the dedicated measurements, $0.184$ of the crystals-retracted loss, is also not directly comparable with production: the scans and the Sobol map of Section~\ref{sec:expdata} ran at low intensity, most likely with a smaller emittance than the production beam, and a smaller emittance deepens the simulated dip (Section~\ref{sec:expdata:emitzs}).

\paragraph{MFBO methodology.} The surrogate is pre-trained on a prior dataset, the cost model understates the cost of tracking, the discrete coordinates are handled approximately, and manufacturing tolerances are not in the GP noise model, so the surface is an ideal-array prediction (Appendix~\ref{app:mfbo}).

\paragraph{Simulation/machine matching.} The agreement is a fit to the two one-dimensional scans, quoted over the dip window. The matched emittance is where the model most resembles the machine, and whether $3~\mu$m is physically defensible for this beam remains open. Every configuration of the global match and of the emittance map shares one captured initial distribution, so comparisons between them are usable while the absolute loss of any one of them carries the $0.041$ capture term of Table~\ref{tab:uncertainty}.

\paragraph{Feedback controller.} On the plateau the dither stays below the deadband, so the controller reacts only once a drift has raised the smoothed loss to about $0.32$, some $0.04$ above its median; that threshold and the $10~\mu$m / $10~\mu$rad quantisation set the floor on the steady-state accuracy and on the smallest drift corrected before it accumulates.

\paragraph{Magnetic hysteresis.} The optics drift of the SPS main magnets~\cite{Velotti_2024_IPAC} is still present upstream of the goniometer; the array's acceptance and the controller absorb it locally, and its cause is not addressed here.

\section{Applicability to Other Machines}
\label{sec:applicability}

Nothing in the method is specific to the SPS. Septum shadowing addresses the loss mechanism of every third-integer-resonance slow extraction through an electrostatic septum, and the same principle is in use or under study elsewhere: a diffuser is in operation upstream of the septum at the J-PARC Main Ring~\cite{Muto_2024_IPAC}, where a bent silicon crystal is planned in place of one of the two diffusers after FLUKA simulations showed it halving the septum loss~\cite{Muto_2025_IPAC}, and crystal channeling ahead of the electrostatic septa has been studied for the Fermilab Delivery Ring~\cite{Nagaslaev_2024}. What transfers is the chain demonstrated here, a design optimisation run against the machine's own extraction optics, a tracking model matched to the machine and tested on a measurement its objective never saw, and a model-free controller holding a wide-acceptance array. The quantities that decide whether it is worth doing are the ratio of the septum's effective width to the spiral step, which sets the loss there is to remove, and the beam divergence at the crystal against the array's angular acceptance. The first of these reproduces the loss at the SPS and its order of magnitude at J-PARC, the two machines where both numbers are published: at the SPS a $500~\mu$m effective width~\cite{Goddard_2020_diffuser} over a $15$~mm operational step~\cite{Velotti_2019_PRAB} gives $3.3\%$ against a measured $3.4 \pm 0.7\%$ before shadowing~\cite{Fraser_2019_IPAC}, and at the J-PARC Main Ring about $60~\mu$m over a $20$~mm step gives $0.3\%$ against a measured $\approx 0.5\%$~\cite{Muto_2024_IPAC}.

\section{Conclusions and Outlook}
\label{sec:conclusions}

At its measured optimum the 3-crystal Multi-Volume Reflection Array reaches, and at its best point exceeds, the factor-of-four loss reduction predicted by its MFBO design (Section~\ref{sec:mvra_design}), and through routine production under an extremum-seeking controller (Section~\ref{sec:feedback}) it holds a median $3.6\times$ and a mean $3.44\times$ on the pre-shadowing baseline, a little short of the target set by the activation budget. Relative to the crystals-retracted loss, the design estimated a loss of $0.25$, the best measured point reached $0.184$, and sustained operation sits at $0.30$ (median) to $0.31$ (mean). The dedicated measurements ran at low intensity, most likely with a smaller emittance than in production, which, with the operation on the plateau beside the minimum, may explain part of the gap between the best point and operation. The extraction has run at this level through the 2026 physics programme, a margin accepted in operation for the machine and for the programme at the increased proton-on-target request. The chain of design, deployment, simulation/machine agreement and on-line control shows that the MFBO design pointed to a three-crystal array whose built counterpart, although staggered more widely than designed, met the predicted reduction, and establishes the MVRA as a sustainable extension of the non-local single-crystal concept of~\cite{Velotti_2024_IPAC}.

The remaining margin to the activation target lies in a 4- or 5-crystal upgrade: the $n_\mathrm{cry}$ scan of Figure~\ref{fig:mfbo_surface}(b) projects a factor of ten at $n_\mathrm{cry} = 4$--$5$ for an ideally aligned array, on the surface now tested at $n_\mathrm{cry} = 3$; at those crystal numbers the estimate has a wider error bar and should be confirmed by tracking with more particles before the upgrade is specified. Whether the projection survives realistic alignment is the open question of Section~\ref{sec:mvra_design:results}, and the fourteen-strip experience of Section~\ref{sec:concepts:mvra} suggests it is the binding one. A systematic comparison of the controllers used across the multi-year programme, over the single-TECS, TECA-single plus TECS and MVRA configurations, will be reported separately.

\begin{acknowledgments}
The Multi-Volume Reflection Array was designed, built and validated by the DECRYCE project at CERN, and we thank the crystal-production, bender-design, metrology and test-beam teams for the device and for the characterisation data used in Sections~\ref{sec:concepts} and~\ref{sec:expdata:match}.

The first author used an AI tool, Claude Code (Anthropic), for part of this work. Under his direction it assisted in developing parts of the simulation, matching and analysis software behind Sections~\ref{sec:expdata} and~\ref{sec:activation} and of the controller software of Section~\ref{sec:feedback}, and in drafting, editing and checking the manuscript. All ideas, methods, measurements and interpretations are the authors' own, from the work reported here and the earlier work cited. The authors verified every number in the paper against the underlying data and code and every citation against its source; the figures are data plots produced by scripts, and no image was generated or altered by AI. The authors take full responsibility for the content.
\end{acknowledgments}

\section*{Data availability}

The simulation, design-optimisation and analysis code behind this paper, together with the scripts that regenerate every figure from the data they read, are openly available in the \texttt{sx\_xsuite\_sps} repository~\cite{data_repo}, tagged \texttt{paper-arxiv-v1} for this version; its \texttt{paper/README.md} maps each section of the paper to its code and each figure to its command. The per-cycle beam-loss-monitor and crystal-setpoint records underlying Sections~\ref{sec:expdata:meas} and~\ref{sec:feedback:deployment}, the archived simulation scans of the global match of Section~\ref{sec:expdata:method}, the emittance and septum-width map of Section~\ref{sec:expdata:emitzs} and the capture ensembles of Appendix~\ref{app:budget} are stored at CERN and are available from the authors on request. The extremum-seeking controller of Section~\ref{sec:feedback} is built on the public CERN ML extremum-seeking library~\cite{cernml_ES_lib}; its deployment code, the \texttt{teca-es} repository, is CERN-internal and is likewise available from the authors on request.

\appendix

\section{MFBO methodology limitations}
\label{app:mfbo}

The MFBO campaign has four methodology limitations. First, the low-fidelity surrogate was pre-trained on a prior crystal-shadowing dataset, so the exploration is biased towards the surrogate's training basin. The multi-fidelity GP treats the surrogate as a correlated, lower-fidelity observation of the objective and revises its predictions wherever high-fidelity data contradict them. In this campaign, however, only one saved acquisition was placed at high fidelity, at five crystals, so that correction rests almost entirely on the 100 prior high-fidelity samples, and the posterior is least constrained where they are absent, at a single crystal. Second, the cost model is proportional to the fidelity input, so it charges a low-fidelity evaluation $0.4$ of a high-fidelity one. The real ratio is milliseconds against tens of minutes, many orders of magnitude smaller, so the acquisition treated tracking as only 2.5 times as expensive as the surrogate. The sensitivity of the optimisation outcome to this cost ratio has not been characterised. Third, two of the six search coordinates (the number of crystals $n_\mathrm{cry}$ and the VR sign direction) are intrinsically discrete, while the knowledge-gradient acquisition treats the input space as continuous; we addressed this approximately, with a mixed-feature acquisition optimisation that performs the inner-loop optimisation at each value of the discrete coordinate. Fourth, the manufacturing tolerances on each crystal's bending angle, position, and width are not included in the GP's noise model; the optimum returned by the MFBO surface is therefore an ideal-array prediction, and the realised MVRA performance reflects the convolution of this prediction with the hardware-tolerance distribution. The sim/machine matching campaign of Section~\ref{sec:expdata} quantifies the residual gap.

\section{Simulation model and separatrix-snapshot pipeline}
\label{app:simmodel}

All tracking is run on CPU with Xsuite (xtrack $0.78.0$, xpart $0.21.0$, xcoll $0.5.12$), on a lattice whose horizontal tune sits exactly on the third-integer resonance. On this lattice the horizontal optics are $\beta_x = 64$~m, $\alpha_x = -1.74$ at TECA and $\beta_x = 94$~m, $\alpha_x = 2.27$ at the ZS entrance, with a horizontal phase advance of $17.56 \times 2\pi$ from TECA to the ZS, whose fractional part, $200^\circ$, sets with the non-linear extraction optics how a deflection at the array maps onto the septum aperture. The ZS is a perfect absorber of a given interception width: a particle whose trajectory enters that band is lost at that point, and scattering off the septum material and the secondary showers it would produce are not simulated. The idealisation makes the width an \emph{effective} one, which is why it is matched to the measurements. The ZS has several candidate widths an order of magnitude apart: the anode wires are $60~\mu$m in diameter in the first two septum units and $100~\mu$m in the remaining three~\cite{Velotti_2019_PRAB}; the effective width that reproduces the measured loss once wire straightness and unit-to-unit alignment are folded in was measured at $500~\mu$m~\cite{Velotti_2019_PRAB,Goddard_2020_diffuser}; and the non-local device was specified against an assumed effective thickness of $400~\mu$m. The width matched here is a fourth quantity, belonging to this model: it counts every trajectory crossing the band as lost, with no grazing-incidence transparency between the wires and no scattered proton continuing into the gap, so it is expected to be, and is found to be, smaller than the geometric envelope. A consequence is that the matched model under-predicts the absolute septum loss: at the operating point the $1.4\%$ of the extracted beam that the array rescues, at a loss ratio of $0.26$ (Section~\ref{sec:expdata:matching}), implies a crystals-retracted wire loss of about $2\%$ of the extracted beam, against the $3.4 \pm 0.7\%$ measured before shadowing~\cite{Fraser_2019_IPAC}. Only loss ratios are used in this paper.

A representative separatrix distribution at the TECA s-location is not trivial to construct. A uniform circular sector in $(x, p_x)$ within the array's angular acceptance gives simulated loss curves whose angular FWHM is too narrow by a factor of five to six. The reason is structural: the beam arriving on the separatrix during third-integer extraction has the angular spread of the separatrix arms, a triangular structure in $(x, p_x)$ set by the sextupole resonance driving terms and the chromaticity. The two-stage \emph{separatrix-snapshot} pipeline therefore tracks, in Stage~A, a population through the full SPS lattice for $\sim 300$ turns with the crystal removed and the ZS aperture in place, recording on every turn the transverse coordinates of every particle reaching the TECA s-location. In Stage~B that snapshot is replayed through a few turns with the array installed and the ZS losses accumulated as in a standard extraction simulation. The snapshot holds one row per proton per turn at the TECA location, about $2.6$ rows per proton near the extraction front, and every row is replayed as an independent particle, with the array in and retracted alike; the loss ratio therefore weights each proton by the number of turns it spends at the front, identically in numerator and denominator. Because a proton's rows share its fate, the row count overstates the statistics, which is one reason the capture term of Table~\ref{tab:uncertainty} dominates. The few-turn replay reproduces the full-physics 300-turn ZS-arrival distribution to within a Kolmogorov--Smirnov statistic of $0.36\%$ on the transverse position and momentum marginals, inside the statistical noise of the simulation. The resonance-crossing dynamics is thus paid once per beam setting, and the crystal-jaw and angle scans run on the inexpensive replay; because the separatrix structure is set by the optics, a fresh Stage~A snapshot is generated for each value of the normalised emittance.

\section{The global match: optimiser, search space and convergence}
\label{app:globalmatch}

The optimiser is Optuna~\cite{Akiba_2019_Optuna} (version $4.2.1$) driving a tree-structured Parzen estimator~\cite{Bergstra_2011_TPE}, run in its multivariate form so that correlations between the eight parameters are modelled, and with the constant-liar heuristic enabled so that concurrent workers do not repeatedly propose the same point. The first $60$ trials are drawn quasi-randomly to seed the estimator, after which the sampler takes over. The eight parameters are searched on the bounded, discretised ranges of Table~\ref{tab:searchspace}; the grid steps are set to the resolution at which the objective is distinguishable from its own seed noise, so the search does not spend trials separating configurations the metric cannot tell apart. The objective minimised is $\min[\max(d_\mathrm{pos}, d_\mathrm{ang}) + 0.03\,(d_\mathrm{pos} + d_\mathrm{ang}),\, 1]$, where $d_\mathrm{pos}$ and $d_\mathrm{ang}$ are the two dip-region deviations: the $\max$ is what forces both axes down, the small sum term breaks ties between configurations whose worse axis is equal, and the cap keeps a degenerate trial from dominating the estimator. A candidate whose position-scan deviation exceeds $0.05$ is pruned before the expensive angle sweep and scored $1 + d_\mathrm{pos}$, so a pruned trial still carries usable ranking information.

\begin{table}[h]
\centering
\caption{The eight jointly matched parameters, with the bounded ranges and grid steps the global optimisation was given. All five leading configurations landed on the lower bound of the emittance range; no configuration reached a bound on any other axis.}
\label{tab:searchspace}
\begin{tabular}{lcc}
\hline
Parameter & Range & Step \\
\hline
Normalised emittance $n_\mathrm{ex}$ & $3.0$ to $8.0~\mu$m & $0.25~\mu$m \\
Effective ZS width & $100$ to $500~\mu$m & $20~\mu$m \\
Transverse offsets (each of 3) & $-100$ to $+100~\mu$m & $5~\mu$m \\
Angular offsets (each of 3) & $-50$ to $+50~\mu$rad & $2.5~\mu$rad \\
\hline
\end{tabular}
\end{table}

\begin{figure}
\includegraphics[width=\columnwidth]{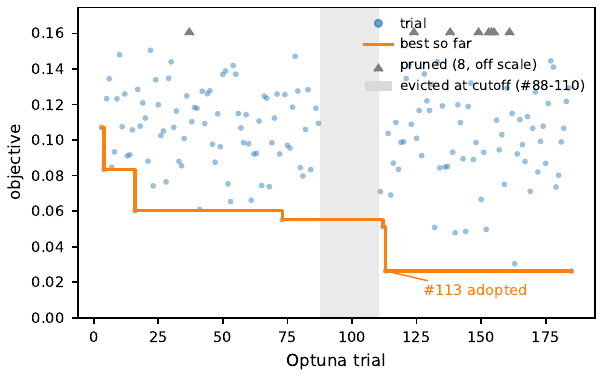}
\caption{Convergence of the global match: the objective of every scored trial against the Optuna trial number, with the running best. Pruned trials sit off scale and are marked along the top; the shaded block is the trial numbers orphaned when the first wave of workers hit the batch-farm time limit and was resubmitted. The adopted configuration, trial 113, is the one that survived the fresh-seed re-scoring; the last step of the running best did not.}
\label{fig:convergence}
\end{figure}

Each trial is considerably more than two scans: a $25$-point position scan at three replay seeds, followed by a bracketed sweep of nine candidate jaw positions each carrying its own $25$-point angle scan at three seeds, which is $750$ replay simulations of $4\times 10^{4}$ particles over six turns, plus a Stage~A capture for any emittance not already cached. The median trial cost was $5.8$~h on four cores. The campaign ran $25$ concurrent workers on the CERN batch farm over $53$~h of wall clock. Of the $186$ trial numbers issued, $159$ ran to completion, $8$ of those pruned at the position gate; the other $27$ were lost to a job-flavour timeout that evicted a wave of $23$ mid-flight, resubmitted under new numbers, and to a few first-submit failures. The total cost was about $3\,400$ core-hours. There is no convergence-based stopping rule: the campaign ran to a fixed trial budget, and the running best is reported as it stands. It plateaued near $0.055$ for roughly forty trials before dropping to $0.026$, which suggests that the budget was adequate; convergence is not claimed (Figure~\ref{fig:convergence}).

The normalised emittance accumulated against the lower bound of $3.0~\mu$m: all five leading configurations sit exactly on it. The optimum is therefore constrained by the box, which motivated the dedicated map of Section~\ref{sec:expdata:emitzs} with the range opened a further micron below.

Three features of the search carry the result. It is global: earlier hand-tuning had brought the angle scan close to agreement while the position scan stayed persistently shallow, and the deficit closed only on a narrow effective septum width, a low emittance and a staggered set of transverse offsets \emph{in combination}, a coupled solution one-parameter scans cannot reach. The metric leaves a bounded lateral offset free, so the offsets that come out, $+2.320$~mm and $+176~\mu$rad, are the fitted calibration between the simulated and machine jaw and goniometer scales, which also absorbs the common part of the per-crystal transverse and angular offsets; the evidence that the angular one is physical is that the angle scan has to cross each strip's channeling condition, an absolute angular fiducial. And a candidate is scored on the larger of its two deviations (Section~\ref{sec:expdata:method}).

\section{How each term of the uncertainty budget was measured}
\label{app:budget}

The terms of Table~\ref{tab:uncertainty} were measured on the simulated loss at the alignment optimum, normalised to the crystal-out loss. The first row, the spread across the three replay seeds of a single run, was read node by node on the emittance and septum-width map of Section~\ref{sec:expdata:emitzs}, where it ranges from $0.005$ to $0.03$; it is the error bar drawn on Figure~\ref{fig:emit_zs_cut}.

The largest term is the initial distribution itself. Four Stage~A captures of the same beam at the same working point, differing only in the seed of the capture, give losses of $0.205$, $0.266$, $0.303$ and $0.271$, a standard deviation of $0.041$. The map of Section~\ref{sec:expdata:emitzs} survives that only because its emittance ladder was captured at a single common seed, so the captures are correlated and the residual row-to-row scatter is $0.018$, some $0.43$ of the independent value; an emittance trend built from independently seeded captures would not have been resolvable at all.

A second, larger measurement exists at a different point of the same configuration. $356$ independent captures of $5\times 10^{4}$ particles over $300$ turns, each replayed five times over six turns at the fixed operating point (jaw $-39.625$~mm, the $-2730~\mu$rad plateau; the same jobs delivered the twelve-turn extracted-beam pair of Section~\ref{sec:expdata:matching}), give a loss of $0.264 \pm 0.029$ (standard deviation; standard error $0.0015$) of the crystals-retracted level, with a $16$th to $84$th percentile range of $0.234$ to $0.292$. The two observables differ, an argmin over nine jaw positions and $25$ angles against a fixed point, and the fixed-point scatter is the smaller, as it should be. The campaign's own capture, read at the same fixed point on the $5~\mu$rad fine scan, gives $0.247 \pm 0.019$ over six replay seeds, $0.6$ standard deviations below the ensemble mean, so it is an ordinary member of the ensemble; on the argmin observable it sits $1.4$ standard deviations below the mean of the four.

The run-to-run term is a caution about seeded replays in general. Nodes deliberately recomputed at an identical configuration reproduce every deterministic quantity exactly, including the crystal-out reference to the individual count, which is consistent with a retracted-crystal replay consuming no random numbers. Only the crystal-in loss moves, by $0.008$ on average and $0.028$ at worst. The replay seed therefore does not by itself fix the scattering stream: that depends on the execution history of the process, so a worker that reuses one tracking engine across several evaluations and a worker that builds a fresh one do not draw the same numbers from the same seed. This is a reproducibility limit of the seeded replay, not physics, and it is of the same size as the replay-seed spread of Table~\ref{tab:uncertainty}.

\section{Estimators chosen by measuring the alternatives}
\label{app:estimators}

The angular working point is read from the raw loss minimum of a scan fine enough to resolve the feature. A $25~\mu$rad grid, and later a $42~\mu$rad one, both stepped over a structure about $15~\mu$rad wide and returned the shoulder beside it (Section~\ref{sec:concepts:nonlocal}). Whether the minimum or the centre of the surrounding basin is the better estimator depends on the shape: for a flat basin the centre is right, because the raw minimum wanders inside it under the scattering noise, while for the narrow feature of Section~\ref{sec:expdata:match} the minimum is right, and the two are distinguished by whether the seeds agree on which point wins.

The jaw position at which each map node is evaluated was chosen the same way. Three conventions were computed at every node and scored by how smooth the resulting map is in emittance. Taking the jaw position the amorphous position scan selects looks like the controlled choice and is by far the worst, because the minimum of a shallow $25$-point scan wanders over more than a millimetre while the useful range is half of that. The convention used, the deepest loss over the node's own jaw-position sweep, was preferred over the jaw position the match metric selects because it is independent of that metric, which keeps the loss map and the agreement map two independent readings of one grid. The residuals about a smooth trend in emittance are $0.015$, $0.017$ and $0.143$ for the three, in that order.

The third is the lateral offset between the simulated and measured scans, left free within a bounded interval (Section~\ref{sec:expdata:method}).


\end{document}